\documentclass[12pt,preprint]{aastex}

\newcommand{\acr}{\beta_{\rm cr}}
\newcommand{\BP}{Ballesteros-Paredes}
\newcommand{\gamef}{\gamma_{\rm e}}
\newcommand{\kms}{{\rm ~km~s}^{-1}}
\newcommand{\lc}{l_{\rm c}}
\newcommand{\Mr}{M_{\rm 1,r}}
\newcommand{\pcc}{{\rm ~cm}^{-3}}
\newcommand{\psc}{{\rm ~cm}^{-2}}
\newcommand{\Peq}{P_{\rm eq}}
\newcommand{\tc}{\tau_{\rm c}}
\newcommand{\Teq}{T_{\rm eq}}
\newcommand{\uu}{{\bf v}}
\newcommand{\vf}{v_{\rm f}}
\newcommand{\VS}{V\'azquez-Semadeni}
\newcommand{\vs}{v_{\rm s}}

\slugcomment{Submitted to ApJ}

\shorttitle{Molecular Cloud and CNM Sheet Formation}
\shortauthors{\VS\ et al.}

\begin{document}


\title{Molecular cloud evolution. I. Molecular cloud and thin CNM sheet
formation} 


\author{Enrique \VS$^1$, Dongsu Ryu$^2$, Thierry Passot$^3$, Ricardo
F.\ Gonz\'alez$^1$ and Adriana Gazol$^1$}
\affil{$^1$Centro de Radioastronom\'ia y Astrof\'isica (CRyA), UNAM,
Apdo. Postal 72-3 (Xangari), Morelia, Michoac\'an 58089, M\'exico}
\email{e.vazquez, rf.gonzalez,a.gazol@astrosmo.unam.mx}
\affil{$^2$ Dept. of Astronomy \& Space Science, Chungnam National
University, Daejeon 305-764, Korea}
\email{ryu@canopus.cnu.ac.kr}
\affil{$^3$CNRS, Observatoire de la C\^ote d'Azur, B.P. 4229,
06304, Nice, C\'edex 4, France}
\email{Thierry.PASSOT@obs-nice.fr}


\begin{abstract}

We analyze the scenario of molecular cloud formation by large-scale
supersonic compressions in the diffuse warm neutral medium (WNM). During
the early stages of this process, a shocked layer forms, and within it,
a thin cold layer. An analytical model and high-resolution 1D simulations
predict the thermodynamic conditions in the cold layer. After $\sim 1$
Myr of evolution, the layer has column density $\sim 2.5 \times 10^{19} \psc$,
thickness $\sim 0.03$ pc, temperature $\sim 25$ K and pressure $\sim
6650$ K $\pcc$. These conditions are strongly reminiscent of
those recently reported by Heiles and coworkers for cold neutral medium
sheets. In the 
1D simulations, the inflows into the sheets produce line profiles with a
central line of width $\sim 0.5 \kms$ and broad wings of width $\sim 1
\kms$. Being the result of the inflows rather than of turbulence, these
linewidths do not imply 
excessively short lifetimes for the sheets. At later times, 3D numerical
simulations show that the cold layer undergoes a dynamical
instability that causes it to develop turbulent motions and to increase
its thickness, until it becomes a fully three-dimensional turbulent cloud.
The destabilization mechanism appears to be the nonlinear thin
shell instability, triggered by the thermal instability-induced
density contrast. In 
the simulations, fully developed turbulence arises on 
times ranging from $\sim 7.5$ Myr for inflow Mach number $\Mr = 2.4$ to
$> 80$ Myr for 
$\Mr = 1.03$. Due to the limitations of the simulations, these numbers
should be considered upper limits. In the
turbulent regime, the highest-density gas (HDG, $n 
> 100 \pcc$) is always overpressured with respect to the mean WNM
pressure by factors 1.5--4, even though we do not include
self-gravity. The intermediate-density gas (IDG, $10 < n [{\rm
cm}^{-3}] < 100$) 
has a significant pressure scatter
that increases with $\Mr$, so that at $\Mr = 2.4$, a significant
fraction of the IDG is at a higher pressure than the HDG. The ratio of
internal to kinetic energy 
density varies from the inflow to the IDG and the HDG, and increases
with density in the most turbulent runs. Our results suggest that the
turbulence and at least part of the excess pressure in molecular clouds
can be generated by the compressive process that forms the clouds themselves,
and that thin CNM sheets may be formed transiently by this mechanism,
when the compressions are only weakly supersonic.

\end{abstract}


\section{Introduction} \label{sec:intro}

Molecular cloud complexes (or ``giant molecular clouds'', GMCs) are
some of the most studied objects in the interstellar medium (ISM) of
the Galaxy. Yet, their formation mechanism and the origin of their
physical conditions remain uncertain \citep[see, e.g., ][]{Elm91,BW99}.  In
particular, GMCs are known to have masses much larger than their
thermal Jeans masses \citep[ sec. 7.2]{ZP74} (i.e., thermal energies
much smaller than their absolute gravitational energy), but comparable
gravitational, magnetic, and turbulent kinetic energies
\citep{Larson81, MG88, Crutcher99, Crutcher04, Bourke01}. This
similarity has traditionally been interpreted as indicative of
approximate virial equilibrium, and of rough stability and longevity
of the clouds \citep[see, e.g., ][]{McKee_etal93,BW99}. In this
picture, the fact that molecular clouds have thermal pressures
exceeding that of the general ISM by roughly one order of magnitude
\citep{Blitz91,MK04} is interpreted as a consequence of the fact that
they are strongly self-gravitating. Because of their
self-gravitating and overpressured nature, molecular clouds were left
off global ISM models based on thermal pressure equilibrium, such as
those by \citet{FGH69} and \citet{MO77}. 

Recent work suggests instead that
molecular clouds and their substructure may actually be transient
features in their respective environments. In this picture, the clouds
and their substructure are undergoing secular dynamical 
evolution, being assembled by large-scale supersonic compressions in
the atomic medium on a crossing time and becoming self-gravitating in
the process \citep{Elm93,Elm00,VPP96,BVS99,BHV99,HBB01, PAL01, Hart03,
Berg_etal04,VKSB05}. 
Specifically, \citet{HBB01} suggested that the accumulation process of
atomic gas may last a few tens of megayears until the gas finally becomes
molecular, self-gravitating and supercritical at roughly the same time
\citep[see also][]{FC86}. Observations of atomic inflows surrounding
molecular gas support this scenario \citep{BHV99, Brunt03}.

In this scenario, the excess pressure in molecular clouds may arise
from the fact that they have been assembled by motions whose ram
pressures are comparable or larger than the local thermal pressure, so
that both the excess pressure and the 
self-gravitating nature of the clouds may originate from the
compression that forms the cloud, and need not be an indication of any
sort of equilibrium \citep{Maloney90,BV97,BVS99,Clark_etal05}. 
Additionally, it has been suggested by  
\citet{VBK03} that a substantial fraction of the internal turbulence
of the clouds may originate from thin-shell instabilities
in the compressed gas, rather than from local energy injection from the
clouds' own stellar products \citep[see also][]{HBB01}. This would
explain the fact that even clouds with no apparent stellar content
\citep[see, e.g., ][]{MT85} have levels of turbulence 
comparable to clouds with healthy star-forming activity. 

Since the typical velocity dispersion in the warm medium (which we will
denote generically as WNM, as we
will not distinguish between neutral and ionized warm components)
is $\sim 10$ km s$^{-1}$ \citep{KH87,HT03}, the flow is transonic
(i.e., with rms Mach number $\sim 1$). Under these conditions,
transonic compressions in the WNM may render the gas unstable to
two main types of instabilities. On one hand, there is the thermal
instability (TI), originally studied by \citet{Field65} \citep[see also
the reviews by][]{Meerson96,VS_etal03}, and by
\citet{McCray_etal75} in the context of radiative post-shock cooling
regions, in which it causes the formation of thin cold sheets
parallel to the shock front. The formation of cold neutral medium
(CNM) structures induced by compressions in the WNM has been more
recently studied analytically and numerically by \citet{HP99,HP00} and
\citet{AH05}. 

On the other hand, there are bending mode instabilities, which cause
rippling of the compressed layer, and, eventually, fully developed
turbulence. The nonlinear theory was laid down by \citet{Vish94} in
the isothermal case, who referred to the instability as nonlinear
thin-shell instability (NTSI), a name which we use throughout this paper.
More recently, \citet{Pittard_etal05} have studied
the stability of cooling shocks and its dependence on the Mach number
and the final post-shock temperature, finding that lower final
temperatures cause greater overstability. Numerically, several groups
have studied the problem in a variety of contexts, such as the
interiors of molecular clouds at 
sub-pc scales \citep{Hunter_etal86}, stellar wind interactions
\citep{Stevens_etal92}, cloud collisions \citep{KW98}, and in a
general fashion \citep{WF98,WF00}. In particular, the simulations of the
latter authors show that turbulence is generated and maintained
for the entire duration of the inflowing motions. 

In the context of the thermally
bistable atomic medium, a two-dimensional study of converging flows
in the diffuse atomic medium has been recently presented by
\citet{AH05}, but focusing on the competition between externally
imposed turbulence and the tendency of TI to produce two-phase
structure, rather than in the production of turbulence by the
process. In a related context, \citet{KI02} and \citet{IK04} have studied the
production of supersonic turbulence behind a propagating shock wave in
the warm neutral ISM, showing that cold, dense cloudlets are formed,
with bulk velocities that are supersonic with respect to their internal
temperatures, proposing this mechanism as the origin of molecular
cloud turbulence. However, these authors attributed the development of
turbulence to TI alone without indicating the
precise mechanism at play, and focused on the
structure behind a single traveling shock wave, rather than on the
structure in shocked compressed layers between converging
flows. Instead, large molecular masses typical of GMCs probably
require focused, large-scale compressive motions that may last several
tens of megayears, as in the scenario proposed by \citet{HBB01}. In
this case, the turbulence generation might last for as long as the
accumulation (compression) lasts, and the turbulence in molecular
clouds could be considered as driven, rather than decaying. A
preliminary study of the development of the turbulence in these
circumstances has been recently presented by \citet{Heitsch_etal05}, who
also investigated the masses and possible self-gravitating nature of the
clumps formed by the induced turbulence.

The advanced stages of molecular cloud evolution are also poorly understood. 
Traditionally, the greatest concern about molecular clouds has been
how to support them against self-gravity, and it was speculated that
turbulence could be prevented from decaying if it consisted primarily
of MHD waves \citep{AM75}, although numerical
work strongly suggests that MHD turbulence decays just as fast as
hydrodynamic turbulence \citep{ML_etal98, SOG98, ML99, PN99,
LB05}. Clouds were assumed to be finally dispersed by the kinetic energy 
injection from their stellar products \citep[e.g.,][ sec.\ 5]{Elm91}

Instead, in the secular-evolution scenario cloud support may 
not be a concern at all. Clouds start as atomic entities with negligible
self-gravity that increase their mean density as they are compressed, so
that their self-gravity also increases in the process. When they
finally become strongly self-gravitating, collapse
\emph{does} occur, albeit in a localized fashion that involves a small
fraction of the total cloud mass, due to the action of their internal
turbulence, which forms nonlinear density fluctuations that may
collapse themselves on times much shorter than the parent cloud's
free-fall time \citep{Sasao73, Elm93, Padoan95, VPP96, BVS99, KHM00, HMK01,
PN02, Li_etal04, LN04, NL05, VKSB05, CB05}. The mass that does not
collapse sees its density reduced, and may possibly be dispersed as soon as
the external compression ends. Indeed, even in simulations of decaying,
non-magnetic self-gravitating turbulence, the star formation
efficiency is smaller than unity \citep[e.g., ][]{BBB03}, and some of
the cloud's mass is dispersed away from the simulation box
\citep{CB04,Clark_etal05}.

The present paper is the first in a series that analyzes
quantitatively the turbulent scenario of molecular
cloud evolution, assuming that they are formed by large-scale
compressions in the warm diffuse medium. Here
we focus on the formation stages, investigating the physical
conditions that may be produced in the compressed layer as a function
of the Mach number of the converging streams, and so we neglect the
chemistry and self-gravity of the gas and use moderate-resolution
simulations to test the 
order-of-magnitude estimates that can be made from simple
considerations, the triggering of thin-shell instabilities, and the
statistics of the physical conditions in the turbulent gas. 

We also report on the formation of transient, thin sheets of cold gas
during the initial stages of the cloud formation process. These sheets
turn out to closely match the properties of cold neutral medium (CNM)
sheets recently observed by Heiles and collaborators
\citep{HT03, Heiles04}. These authors have reported on the existence of
extremely thin CNM sheets, with thicknesses $\sim 0.05$ pc, column
densities, $\sim 0.2 \times 10^{20} \psc$, temperatures $\sim 20$ K and
linewidths $\sim 1~\kms$, and have argued that these properties are
difficult to understand in terms of turbulent clouds. We will argue that
objects with these properties are naturally formed as part of the cloud
formation process, in cases of low-Mach number compressions.

The plan of the paper is as follows: In \S
\ref{sec:basic_phys} we describe the basic physics of the pre-turbulent
stages of the system, and then in \S \ref{sec:num_sim}
we present numerical simulations that confirm the estimates and
follow the nonlinear stages in which the compressed and cooled layers
become turbulent. We describe the numerical method in \S
\ref{sec:num_meth}, relevant resolution considerations in \S
\ref{sec:num_res}, and the results in \S
\ref{sec:results}. In \S \ref{sec:discussion} we discuss our results,
comparing them with observational data and previous work. Finally, in 
\ref{sec:concl} we give a summary and some conclusions.

\section{The physical system} \label{sec:basic_phys}

\subsection{Governing equations} \label{sec:eq}

We consider the atomic medium in the ISM, whose flow and thermal
conditions are governed by the equations 
\begin{eqnarray}
&&\frac{\partial \rho}{\partial t} + \nabla \cdot (\rho \uu) = 0
\label{eq:cont}\\ 
&&\frac{\partial (\rho \uu)}{\partial t} + \nabla \cdot (\rho \uu \uu) = -
\nabla P \label{eq:mom} \\
&&\frac{\partial E}{\partial t} + \nabla \cdot \left[(E + P) \uu\right]
= \rho \left(\Gamma - \rho \Lambda(T)\right), \label{eq:int_en}
\end{eqnarray}
where $\rho$ is the gas density, \uu\ is the fluid velocity,
$E=P/(\gamma -1) + \rho |\uu|^2/2$ is the total
energy per unit volume, $P$ is the thermal
pressure, $e=P/[\rho(\gamma-1)]$ is the internal energy per unit mass
$T = e/c_{\rm V}$ is the temperature, $c_{\rm V} = 
[\gamma(\gamma-1)]^{-1}$ is the specific
heat at constant volume, $\gamma=5/3$ is the heat capacity ratio of the
gas, $\Gamma$ is the heating rate, and $\rho \Lambda$ is the radiative 
cooling rate. We use a piecewise power-law fit to the cooling function,
based on a fit to the standard thermal-equilibrium (TE) pressure versus
density curve of \citet{Wolfire_etal95}, as decribed in
\citet{SVG02} and \citet{GVK05}. The fit is given by 
\begin{equation}
\Lambda (T)=\left\{ \begin{array}{ll}
                0      &\mbox{$T<15\: K$} \\
          3.42\times 10^{16}T^{2.13} &\mbox{$15\: K\leq T<141\: K$} \\
          9.10\times 10^{18}T        &\mbox{$141\: K\leq T<313\: K$} \\
          1.11\times 10^{20}T^{0.565}&\mbox{$313\: K\leq T<6101\: K$} \\
          2.00\times 10^{8} T^{3.67} &\mbox{$6101\: K\leq T$.}
                        \end{array} \right\}
\label{eq:cooling}
\end{equation}
The background heating is taken as a constant
$\Gamma = 2.51 \times 10^{-26} {\rm erg}\;{\rm s}^{-1} {\rm H}^{-1}$,
where ``H$^{-1}$'' means ``per Hydrogen atom''. This value is roughly
within half an order of magnitude 
of the value of the dominant heating mechanism (photo-electric heating)
reported by \citet{Wolfire_etal95} throughout the range $10^{-2} {\rm
cm}^{-3} \leq n \leq 10^3 {\rm cm}^{-3}$. Note that we have for now
neglected magnetic fields, self-gravity, and heating and cooling
processes adequate for the molecular regime.

\begin{figure}
\epsscale{.75}
\plotone{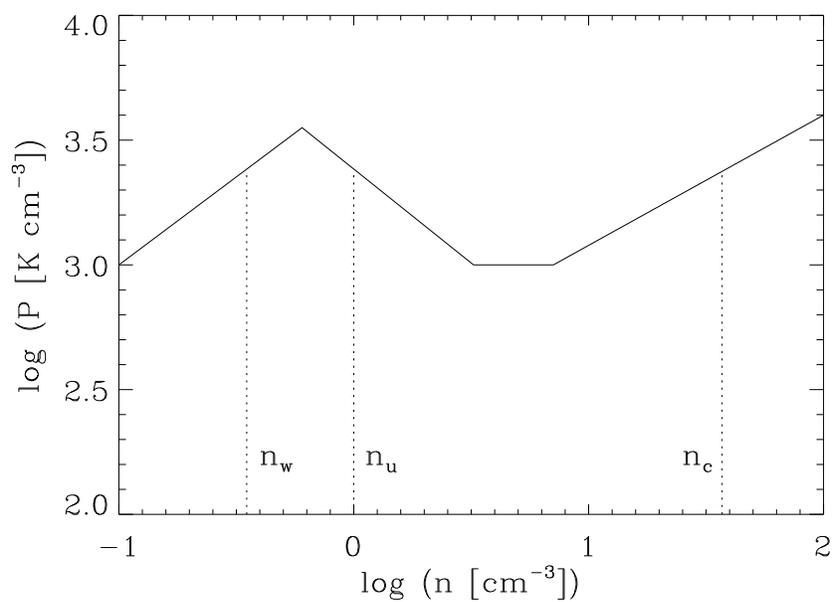}
\caption{Plot of the thermal-equilibrium (TE) value of the pressure
versus the number density implied by the piecewise power-law fit to the
cooling function given by eq.\ (\ref{eq:cooling}) and the assumption of
a constant heating rate. The vertical dotted lines indicate the values
of the density that can coexist in pressure equilibrium, with $n_{\rm w}$ and
$n_{\rm c}$ being stable equilibria (respectively of the warm and cold
phases) and $n_{\rm u}$ being unstable.} 
\label{fig:Peq_vs_rho}
\end{figure}

The condition of thermal equilibrium  between heating and cooling at a
given density defines TE-values of the temperature and thermal pressure,
which we denote by $\Teq(\rho)$ and $\Peq(\rho)$. Figure
\ref{fig:Peq_vs_rho} shows $\Peq$ versus number density $n$ for the
cooling and heating functions defined above. As is well known
\citep{FGH69}, at the mean midplane thermal pressure of the ISM 
\citep[$\sim 2250$ K cm$^{-3}$ or sightly higher,][]{JT01}, the atomic
medium is thermally 
bistable. For our chosen fits to the cooling and heating functions, and
a mean pressure of $2400$ K$\pcc$, a warm diffuse phase ($n_1 \sim 0.35$
cm$^{-3}$, $T \sim 7400$ K) is able to coexist in
pressure equilibrium with a cold dense 
one ($n_2 \sim 37$ cm$^{-3}$, $T \sim 65$ K). A third, unstable phase with
($n_3 \sim 1$ cm$^{-3}$, $T \sim 2400$ K) also corresponds to the same
equilibrium pressure, but is not expected to exist under equilibrium
conditions because it is unstable. The equilibrium values of the
density are also shown in fig.\ \ref{fig:Peq_vs_rho}. Due to our neglect
of molecular-phase cooling and heating, our
$\Peq$-$n$ curve does not correctly represent the approximately isothermal
behavior of molecular gas above densities of a few hundred cm$^{-3}$,
and so we introduce a slight error in the thermodynamic conditions of
the densest gas. We expect to address this shortcoming in subsequent
papers, which will also take into account the chemistry in
the gas and the transition to the molecular phase.

\subsection{Problem description and physical discussion}
\label{sec:setup}

\subsubsection{Analytical model of the early stages} \label{sec:early}

Within the warm medium described above, we consider the collision of two
oppositely-directed gas streams with speeds comparable to the sound
speed, since the velocity dispersion in the warm ISM is known to be
roughly sonic \citep[e.g.][]{KH87, HT03}. This compressive motion may
have a variety of 
origins (the passage of a spiral density wave, or a large-scale
gravitational instability, or general transonic turbulence in the
diffuse medium), whose details are not of our concern here. The
discussion below on the early evolution of the system is inspired by
that of \citet{HP99}, except that in our case we assume that the inflows are
maintained  in time rather than being impulsive, and that our
simulations are three-dimensional.

\begin{figure}
\epsscale{.75}
\plotone{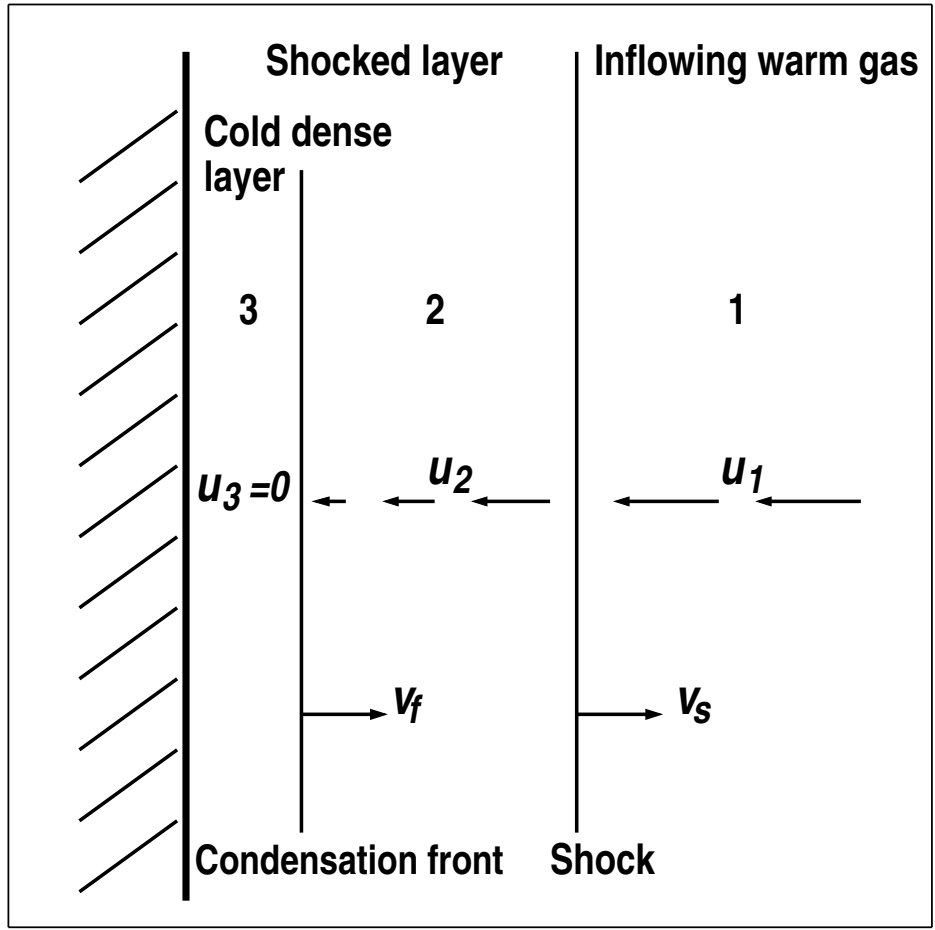}
\caption{Schematic diagram showing the right-hand-side half of the physical
system. The warm diffuse gas, whose physical conditions are labeled with
subindex ``1'', enters from the right, and encounters a 
shock, initially caused by the collision with the opposite stream, shown
in this figure as a wall at the left of the figure. The shock
stops after a time of the order of the cooling time. The immediate
post-shock values of the physical variables are labeled with subindex
``2''. Past the
shock, the flow is subsonic all the way through the wall, and
constitutes the ``shocked layer''. Finally,
very near the wall, at times larger than the cooling time, the gas
undergoes thermal instability and condenses into a thin layer of cold
gas, whose physical variables are labeled with subindex
``3''. Velocities in the rest frame of the 
figure are denoted by $v$, while velocities in the frame of the shock
are denoted by $u$. The entire system is symmetric with respect to the
wall.} \label{fig:schematic}
\end{figure}

The colliding streams have the density and thermal pressure conditions
of the warm diffuse phase, forming a shock-bounded slab at the collision
site, as illustrated in fig.\ \ref{fig:schematic}. During the initial
stages the evolution is adiabatic, the bounding 
shocks move outwards from the collision site, and the gas 
inside the shocked slab is heated and driven away from
thermal equilibrium by the shocks. The gas in the slab then has
conditions dictated by the adiabatic jump relations \citep[see,
e.g.,][]{Shu92}. For later use, we transcribe here the jump conditions
for the velocity and the pressure:
\begin{eqnarray}
u_2 &=& \left(\frac{\gamma -1 + \frac{2}{M_1^2}}{\gamma + 1}\right) u_1,
\label{eq:u_jump}\\
%
%
P_2 &=& \left(\frac{1 - \gamma + 2 \gamma M_1^2}{\gamma + 1} \right)
P_1, \label{eq:P_jump}
%
\end{eqnarray}
where the subindex ``1'' denotes the pre-shock (or upstream) quantities,
subindex ``2'' denotes quantities \emph{immediately}  downstream of the shock,
and the symbol $u$ denotes velocities in the frame of reference of the
shock. Velocities in the rest frame are denoted by the symbol $v$.

%

Denoting by $\vs$ the shock speed (in the rest frame) one has
\begin{equation}
u_2 = v_2-\vs \label{eq:u2_vs}
\end{equation}
and
\begin{equation}
u_1 = v_1 - \vs . \label{eq:u1_v1_vs}
\end{equation}
At the time of the shock formation, the velocity $v_2$ is simply
zero and it is thus possible to calculate the initial shock speed by
combining eqs.\ (\ref{eq:u_jump}), (\ref{eq:u2_vs})  and
(\ref{eq:u1_v1_vs}). One  obtains the following quadratic equation for 
$\vs$ as a function of $v_1$
\begin{equation}
2 \vs^2 - \vs v_1 (3 -\gamma) - \left[v_1^2 (\gamma - 1) + 2 c_1^2
\right] =0,
\end{equation}
where $c_1 = \gamma P_1/\rho_1$ is the sound
speed in the (warm diffuse) pre-shock medium. The solution of this
equation, expressed in terms  of the inflow Mach
number in the rest frame $\Mr \equiv v_1/c_1$, is
\begin{equation}
\frac{\vs}{c_1} = \frac{-\Mr (\gamma - 3) \pm \sqrt{{\Mr}^2 \left[
(\gamma - 3)^2 + 8 (\gamma - 1) \right] + 16}}{4}. \label{eq:vs_M1r}
\end{equation}
%
 

As time progresses, the gas in the slab begins to cool
significantly, so that most of the internal energy increase caused
by the shock is radiated away after roughly a cooling time,
which we take here as 
\begin{equation}
\tc \equiv \frac{c_{\rm V} T_2}{\rho \Lambda(\rho_2,T_2) -
\Gamma}, \label{eq:tau_cool}
\end{equation}
where 
$\Lambda(\rho_2,T_2)$ is the cooling function evaluated at the
immediate post-shock conditions. Equation (\ref{eq:tau_cool}) is well
defined in the shock-heated, out-of-TE post-shock gas,
in which $\rho \Lambda(\rho_2,T_2) > \Gamma$. 

At $t \sim \tc$, the flow returns close to its local equilibrium
temperature. 
If the adiabatic-jump value of the immediate post-shock
pressure is higher than the maximum TE pressure 
allowed for the warm gas, then the cooling brings the gas to the cold
branch of the TE curve, creating a thin, dense, cold layer in
the middle of the compressed slab \citep[][ see also Hennebelle \&
P\'erault 1999]{McCray_etal75}. We
interchangeably refer to this dense layer as the ``cold'' or the
``condensed'' layer, and to its boundary as the condensation front.
The strong cooling undergone by the gas as it flows into
the central regions causes the thermal pressure to drop to
values comparable to that of the inflow, the density to
increase, and the outer shock to decelerate. TI exacerbates
this process by creating such a large density contrast
($> 100$) between the inflow and the central layers
that a quasi-stationary situation is reached
in which shocked layer thickness becomes almost constant and
the outer shock is almost at rest. For $t \gtrsim \tc$, the cold
slab has a half-thickness of roughly one 
cooling length $\lc$, given by $\lc \approx v_2 \tc$,
and the velocity field across the shocked layer takes values between
$v_2$ (immediately behind the shock) and zero at the center of the cold
slab. 

The cooling time
($\tc$) and length ($\lc$) as functions of the inflow Mach number are
plotted in fig.\ \ref{fig:cool_time_length}. As we will discuss in \S
\ref{sec:num_meth}, the cooling length constitutes the minimum physical
size that the numerical box must have.


The pressure in the central cold layer can be estimated assuming
that the total pressure, sum of the kinetic and ram pressures, is
constant in the shocked region.  
The immediate post-shock
pressures and velocities are given by the jump conditions, eqs.\
(\ref{eq:u_jump}) and (\ref{eq:P_jump}), 
using an upstream velocity equal to the inflow speed, since at this time
the shock is not moving in the rest frame. 
The central pressure $P_3$ (in what follows, we denote quantities in the
condensed layer by the subindex ``3'') is thus estimated as
\begin{equation}
P_3 = P_2 + \rho_2 v_2^2
\label{eq:P3P2}
\end{equation}
since the velocity is zero at the center of the condensed layer.

\begin{figure}
\epsscale{0.75}
\plotone{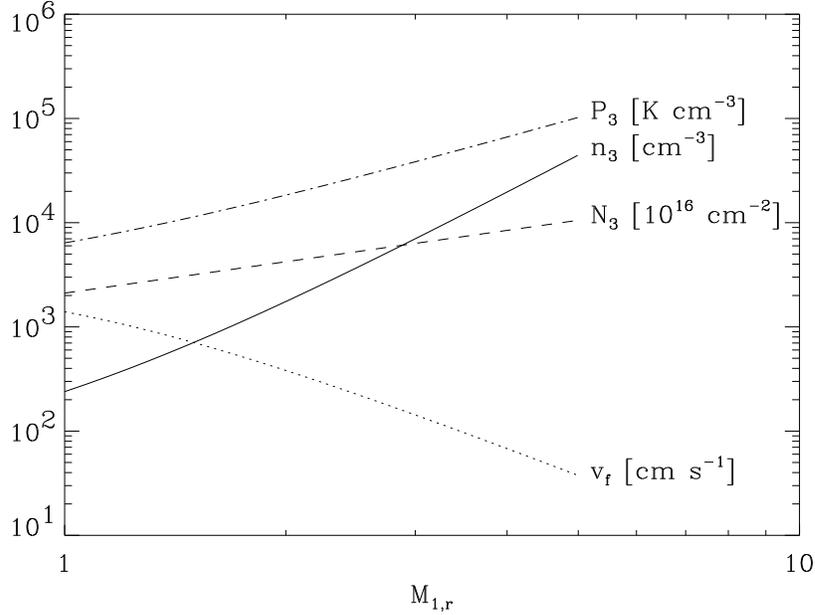}
\caption{Various quantities in the condensed cold layer as a function of
the inflow Mach number $\Mr$, according to the analytical model of
\S \ref{sec:early}: \emph{Solid} line: volume density in the slab
($n_3$). \emph{Dotted} line: outward speed of cold layer
boundary. \emph{Dashed} line: column density through dense
slab after $10^6$ yr. \emph{Dash-dotted} line: Pressure in the cold
layer ($P_3$).} 
\label{fig:ncold_Minfl}
\end{figure}

The density in the condensed layer can then be estimated from the fact
that its constituent gas has undergone a phase transition from the warm
to the cold phase, in which the approach to TE is extremely fast. This
means that the density in the 
cold layer is given by the thermal-equilibrium value at $P_3$.
For a power-law cooling function $\Lambda \propto T^\beta$ and a
constant heating function, it can be
shown that the equilibrium pressure satisfies $\Peq \propto
\rho^{\gamef}$, with $\gamef = 1-1/\beta$ being an effective polytropic
exponent \citep[see, e.g., ][]{VPP96}. Thus, the dense branch of our $\Peq$
vs.\ $\rho$ curve, in which $\beta = 2.13$, is described by the equation 
\begin{equation}
\Peq = 350 \left(\frac{n}{{\rm 1~cm^{-3}}}\right)^{0.53} {\rm K~cm}^{-3}.
\end{equation}
Setting $\Peq = P_3$ 
in the above equation and inverting to solve for the density, we obtain
the dependence of the 
cold layer density $n_3$ on the inflow Mach number $\Mr$, shown as the
solid line in fig.\
\ref{fig:ncold_Minfl}. Note that for molecular gas the
density is not expected to increase as rapidly with $\Mr$, because in
this case the flow is approximately isothermal so that $\gamef \sim 1$
\citep{SVCP98,SS00}. 

If the problem were strictly one-dimensional, then the condensed layer
would simply thicken in time, with its bounding front moving at a speed
$\vf \approx (n_2/n_3) v_2$, an estimate based on mass conservation
from the immediate post-shock region to the cold slab.

This velocity is shown as
the dotted line in fig.\ \ref{fig:ncold_Minfl}. We see that, somewhat
counterintuitively, the front
speed decreases with increasing $\Mr$. This is because
the density contrast between the condensed layer and its surroundings
increases rapidly with $\Mr$, and thus the mass 
entering the condensed layer is compacted very efficiently, causing a
very mild increase in the slab thickness. Finally, the dash-dotted line
in fig.\ \ref{fig:ncold_Minfl} gives the pressure in the 
condensed layer.

The column density through the slab increases with time and is
given by $N_3 = 2 n_3 \vf \Delta t$, where $\Delta t$  is the time since
the beginning of the condensation process, and the factor of 2 comes
from the fact that $\vf\Delta t$ is the half-thickness of the cold
layer. The dashed line in fig.\ 
\ref{fig:ncold_Minfl} shows the column density through the slab after
1 Myr in units of $10^{16}$ cm$^{-2}$, as a function of $\Mr$. We
see that the column density in the cold layer varies relatively slowly
with $\Mr$, with values $2 \times 10^{19}$--$10^{20}$ cm$^{-2}$
at $\Delta t = 1$ Myr. 


The simple model given in this section then predicts the values of the
physical variables in the cold layer. In \S \ref{sec:mod_sim_comp} we
compare its predictions with the results of 1D high-resolution numerical
simulations of the process.

\subsubsection{Late stages} \label{sec:late}

The description of the late stages of the evolution requires numerical
simulations, which we present in \S \ref{sec:num_sim}. Here we just
give some general discussion and expectations. 

Shock compressed layers
are known to be \emph{nonlinearly} unstable in the isothermal case
\citep{Vish94}, meaning that large enough inflow Mach numbers are needed
to trigger the instability. Specifically, displacements of the
compressed layer comparable to its thickness are necessary to trigger
the instability. This would seem to create a difficulty for
destabilizing the compressed layers in the WNM, given the transonic
nature of the flow. 

However, cooling apparently helps in bringing down
the threshold for instability, as suggested by the recent results of
\citet{Pittard_etal05}. These authors found that there exists a critical
value $\acr$ of the cooling exponent for the appearance of the
well-known global overstability in radiative shocks \citep{LCS81}, and
that this exponent depends on 
both the upstream Mach number and the ratio $\chi$ of the cold layer
temperature to the pre-shock temperature. The relevant result for our
purposes is that low-Mach number, low-$\chi$ shocks have values of $\acr$
that are comparable to those of high-Mach number shocks with high
$\chi$. That is,
lower cold-layer temperatures bring down the required Mach number for
overstability. Since the temperature ratio between the cold and warm
phases in the atomic medium is $\chi \sim 0.01$,
the compressed layers formed by transonic compressions in this medium
can probably become overstable, producing oscillations that can bend the
layer strongly enough to trigger a nonlinear thin shell-like (NTS-like)
instability even at relatively low
Mach numbers. Indeed, in the simulations reported in \S
\ref{sec:num_sim}, we have always found that the compressed layer
eventually becomes unstable, given enough time and resolution. We have 
found no threshold for suppression of the NTS-like instability in the
range of inflow Mach numbers we have explored.

\section{Numerical simulations} \label{sec:num_sim}

In this section we present moderate-resolution simulations that are
intended mostly as an exploratory tool of parameter space and of the
phenomena described in the previous section. Detailed, high-resolution
simulations using adaptive-mesh and smoothed-particle hydrodynamics
techniques will be presented in future papers to
investigate the details of the small-scale gasesous structures as well
as the star formation that result in this scenario.

\subsection{Numerical method and limitations} \label{sec:num_meth}

We solve the hydrodynamic equations together with the energy
conservation equations using Eulerian hydrodynamics code
based on the total variation diminishing (TVD) scheme \citep{ryu93}. It
is a second-order accurate upwind scheme, which conserves mass, momentum
and energy. We include the heating and cooling functions described in \S
\ref{sec:eq} as source terms after the hydrodynamic step. 
Tests have shown that this procedure is accurate enough, so that we
did not have to implement an``operator splitting'' algorithm (where
the execution of the hydrodynamic and source stages is alternated)
which formally preserves the second order accuracy.


We solve the equations both in one (1D) and three (3D) dimensions. The
1D runs are used for testing the analytical model of \S
\ref{sec:early}. The 3D runs, in which the inflows enter along the $x$
direction, use periodic boundary conditions along the $y$ and $z$
directions and the same resolution in all three directions (200 grid
cells), except for two runs that use a lower resolution (50 grid cells)
in the $z$ direction and open boundary conditions in $y$ and $z$, that
we used at early stages of this work for studying the dependence of the
time for turbulence development on the $x$ and $y$ resolution. 

In the 3D runs, a random component in time and space, of amplitude $0.5
v_0$, is added 
to the inflow speed at each cell of the boundary as a generic way of
triggering the dynamical instabilities of the compressed layer without
the biasing that might be introduced if we used, for example,
fluctuations with a pre-defined power spectrum. Being so
fast and small-scale, most of this component is erased by diffusivity,
so that the velocity inside the box fluctuates only by a few percent. 

No explicit term for heat conduction is included. In the ideal, absolutely
non-conducting case, the fastest-growing mode of TI has
a vanishing length scale \citep{Field65}. Numerically, this would imply
instability at the scale of the grid size, producing numerical
instabilities. In the presence of
thermal conductivity, the smallest unstable length
scale (the ``Field length'') is finite \citep{Field65}, and sensitively
dependent on the local temperature. If the Field length is not resolved,
numerical diffusion still stabilizes the smallest available scales,
producing a ``numerical Field length'' larger than the real one
\citep{GVK05}, even if 
explicit heat conduction is not included, and in this case the scale of
the fastest-growing mode is resolution-dependent \citep{KI04}.
In practice, this means that if the resolution is insufficient, the
size of the structures \emph{formed by TI} is not adequately resolved and
determined. However, in the presence of turbulence, in which
large-scale supersonic motions are the main drivers of density
fluctuations rather than the development of TI, the \emph{statistics} of
the dense gas (pressure and density distributions), which will be our
main focus in this paper, are rather
insensitive to the resolution even in simulations without thermal
conduction \citep{GVK05}. We conclude that, in spite of not including
thermal conduction explicitly, the moderate resolution we have used
should cause an overestimate of the sizes of the structures (``clumps'')
formed, but should have no serious effect on the density and pressure
statistics of the dense gas we discuss below.


Aware of the above limitations, we adopt the following convenient set of
units for the simulations: $n_0 = 1$ cm$^{-3}$, 
$T_0 = 10^4$ K, $v_0 = c_0 = 11.74$ km s$^{-1}$, where $c_0$ is the
adiabatic sound speed at $T_0$. All simulations start with inflows at
$T=7100$ K and $n=0.338$ cm$^{-3}$, which correspond to the equilibrium
warm phase at a pressure $P=2400$ K cm$^{-3}$. The inflow's velocity,
given in terms of its Mach number with respect to its
sound speed of 9.89 km s$^{-1}$, is varied between $M=1.03$ to
$M=2.4$. 

\begin{figure}
\epsscale{.75}
\plotone{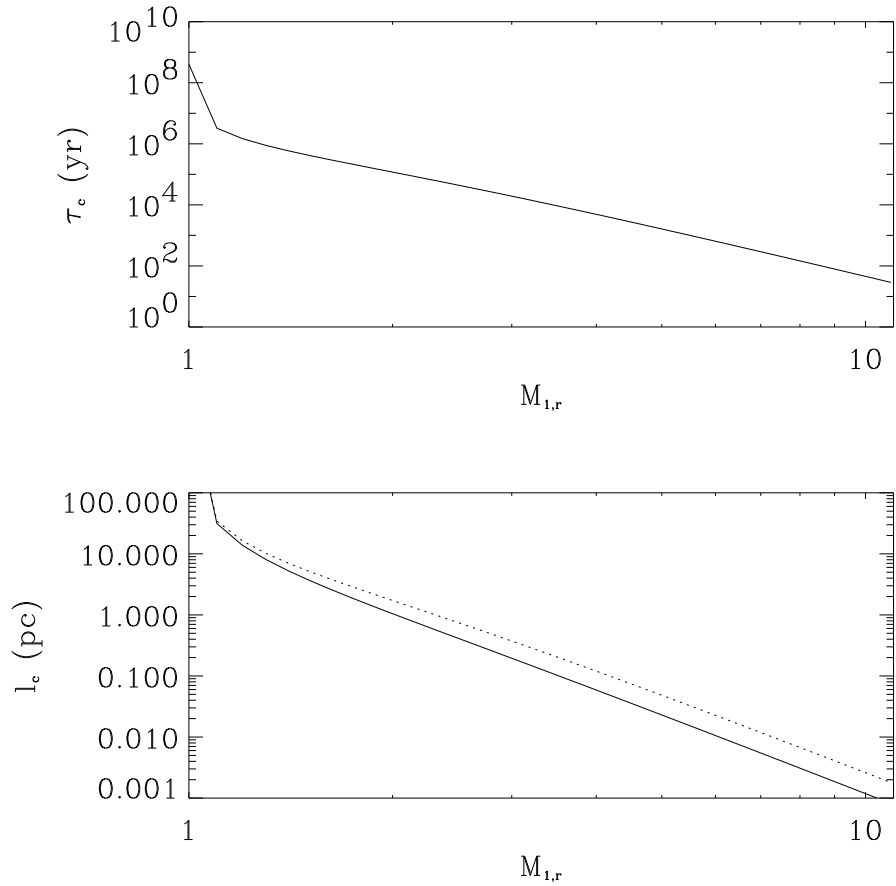}
\caption{\emph{Top:} Cooling time given by eq.\ (\ref{eq:tau_cool}) as a
function of the inflow Mach number in the rest frame,
$\Mr$. \emph{Bottom:} Associated cooling length, computed using the
sound speed (\emph{solid line}) and the inflow speed (\emph{dotted line}).} 
\label{fig:cool_time_length}
\end{figure}

The physical size of the box deserves some special discussion. In \S
\ref{sec:early} we have noted that during the very early stages of the
system's evolution, the shocks move outwards, stopping after roughly one
cooling time, and having traversed one cooling length. In order to
capture the full dynamics of the problem, we must
make sure that the box has a large enough physical size that the shocks
do not leave the simulation before stopping. 
We have found that actual cooling
lengths are larger by factors of $\sim 4$ than indicated by fig.\
\ref{fig:cool_time_length}. Moreover, at late stages of 
the evolution, the slab becomes heavily distorted and thickens
considerably \citep[see also, e.g., ][]{WF00}. So we use box sizes
significantly larger than the cooling length given in fig.\
\ref{fig:cool_time_length}. As seen in this figure, the cooling length is a 
rapidly decreasing function of the inflow Mach number $\Mr$ in the
range we have considered, and so
simulations with smaller $\Mr$ require larger box lengths. Finally, the
physical box size implies a 
physical time unit given by $t_0 = L_0/v_0$, 

We take as a
fiducial simulation one with $\Mr = 2.4$ and a box length $L_0= 16$ pc. 
The coefficients of the cooling function for the fiducial case in code
units are 561.641 for 
15 K $\leq T<141$ K,  4.6178 for 141 K $\leq T < 313$ K, 1.0244 for 313
K $\leq 6101$ K, and 4.7432 for $T > 6101$ K.
Other box sizes (and derived time units) are obtained by simply
multiplying the cooling and heating coefficients by the same factor as
that for the box size, because the only simulation parameter on which
the cooling rates depend is the time unit. 
Table \ref{tab:runs} summarizes the main parameters of the runs we
consider below, in which  the runs are denoted mnemonically by their
inflow Mach number $\Mr$ and their box size $L_0$ as Mn.nLnn, where
'n' is an integer. We denote 1D runs by the suffix ``1D'',
high-resolution versions of another run by the suffix ``hr'', and
simulations with lower resolution in the $z$ direction,
by the suffix ``LZ'', for ``low-$z$''.

\subsection{Resolution considerations} \label{sec:num_res}

\begin{figure}
\epsscale{1.}
\plottwo{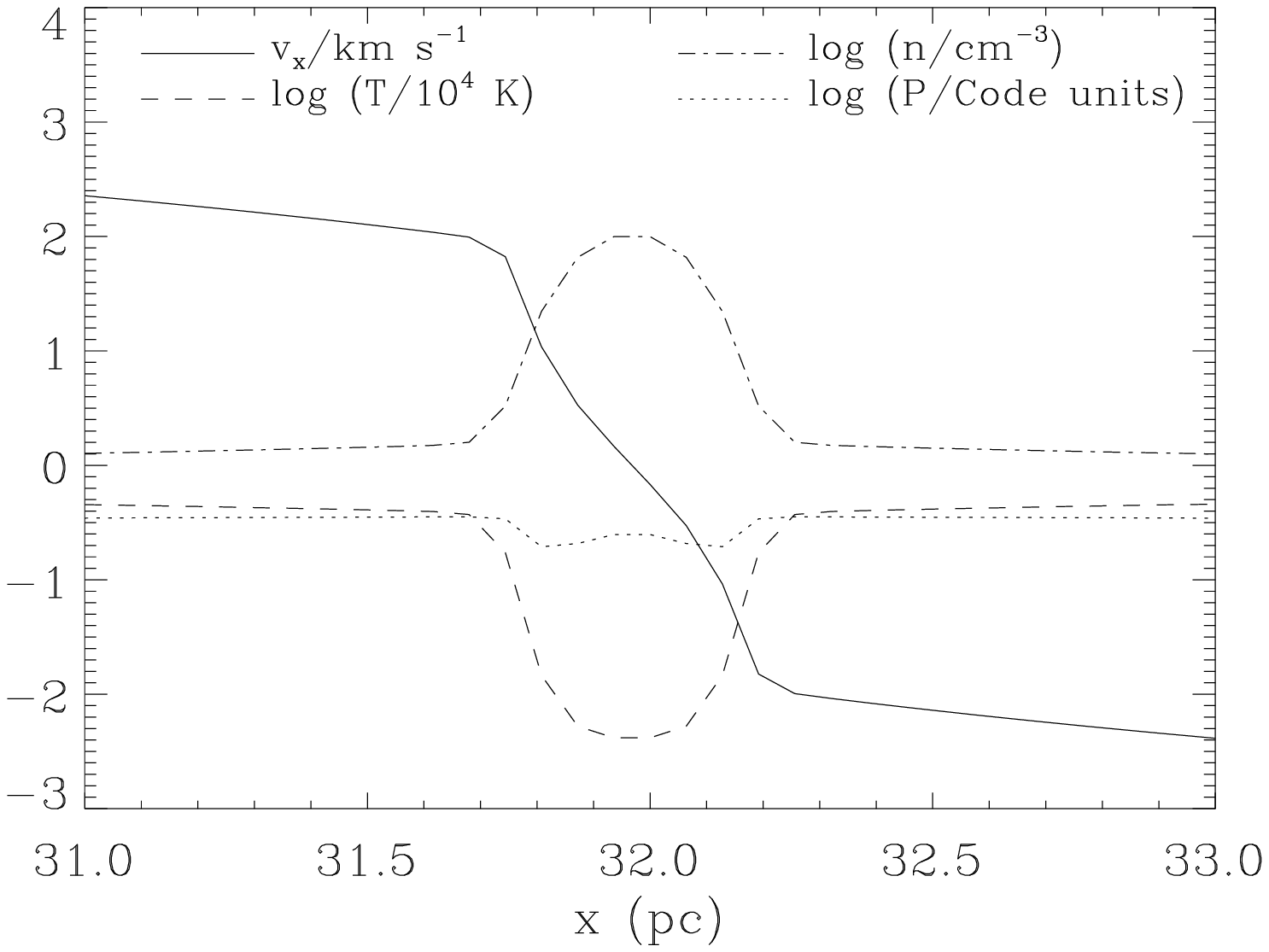}{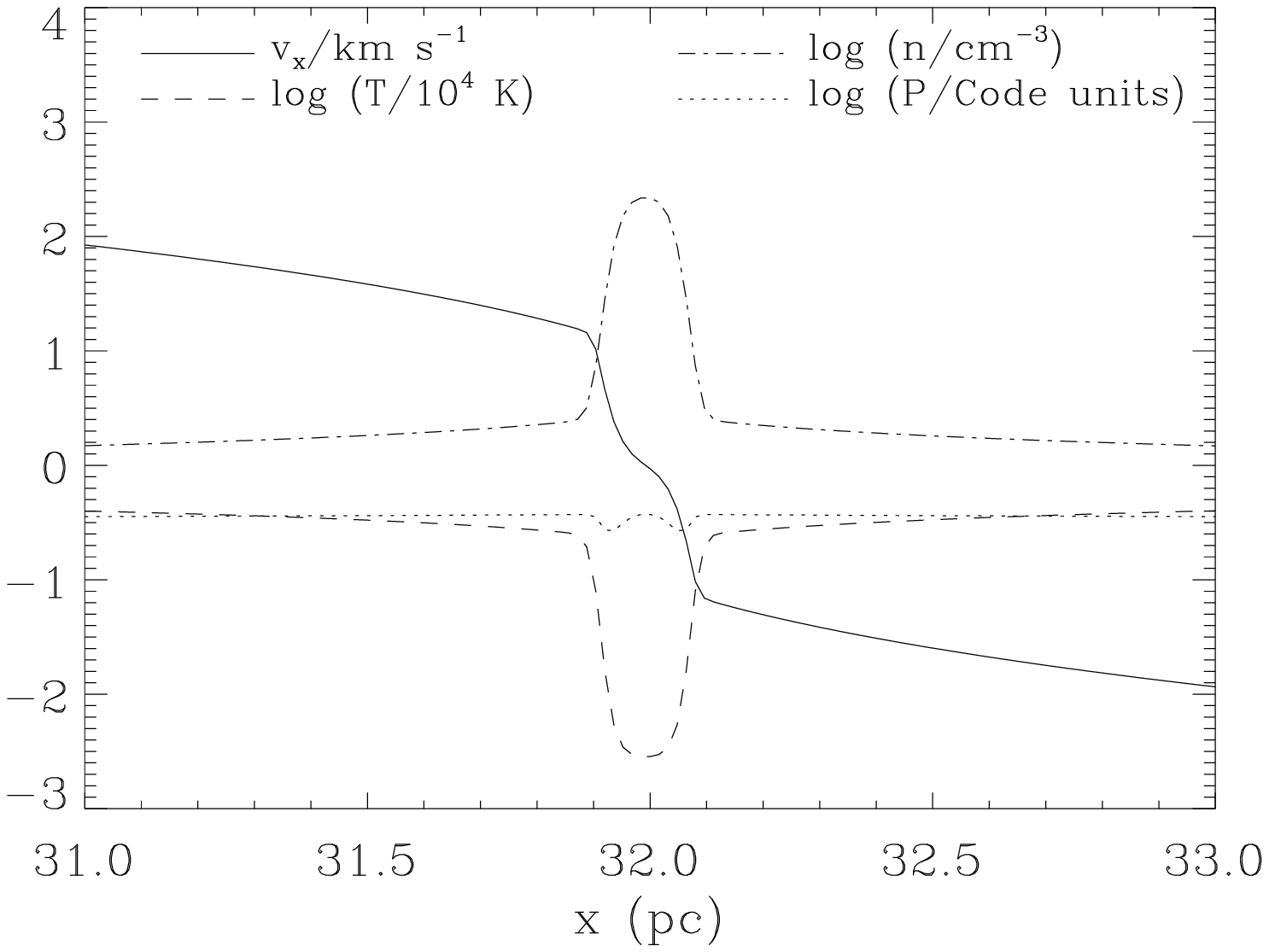}
\caption{Profiles of various physical quantities along the central
regions of runs M1.03L64-1D (\emph{left}) and M1.03L64-1D-hr
(\emph{right}) at $t=5.33$ Myr. Shown are the $x$-velocity (\emph{solid} line),
the density (\emph{dash-dotted} line), 
temperature (\emph{dashed} line), and pressure (\emph{dotted} line).
The latter is given in code units, in which $P=2400$ K$\pcc$
corresponds to a value of 0.144. The cold layer
thickness, density and pressure are seen to depend on the resolution at
the early stages of evolution.} 
\label{fig:Mach0.96}
\end{figure}

Before we discuss the results of the numerical simulations, it is
important to discuss the resolution limitations of the simulations we
present here and their effects, in order to be able to usefully
interpret their results. 
The cold layer is estimated
to be very thin, $\sim 2 \vf \Delta t = 0.029$ pc for $\Mr =
1$, for which $\vf = 0.014 \kms$, when taking $\Delta t= 1$
Myr. Moreover, we have 
seen (fig.\ \ref{fig:ncold_Minfl}) that the front speed $\vf$ is a decreasing
function of $\Mr$, and therefore the cold layer is thinner for higher
$\Mr$. This means that only run M1.03L64-1Dhr should barely
resolve (with $\sim 2$ grid cells) the cold layer at this time. 
In practice the layer is not quite resolved even at this time, since
apparently at least 8--9 zones are necessary to resolve the layer (see
below). 
This implies that the thickness of the cold layer during the early, 
non-turbulent stages depends  on the resolution, and is exaggerated, in
some cases grossly, by our simulations. Figure \ref{fig:Mach0.96} shows
profiles along the $x$-axis at half the $y$ and $z$ extensions of
the $x$-velocity (\emph{solid} line), density (\emph{dash-dotted} line),
temperature (\emph{dahed} line) and pressure (\emph{dotted} line) for
runs M1.03L64-1D (\emph{left panel}) and M1.03L64-1D-hr (\emph{right panel}) at
$t=5.33$ Myr. These runs differ only in the resolution used. It is clear
that the cold layer is thinner in the high-resolution case.

\begin{figure}
\epsscale{1.}
\plottwo{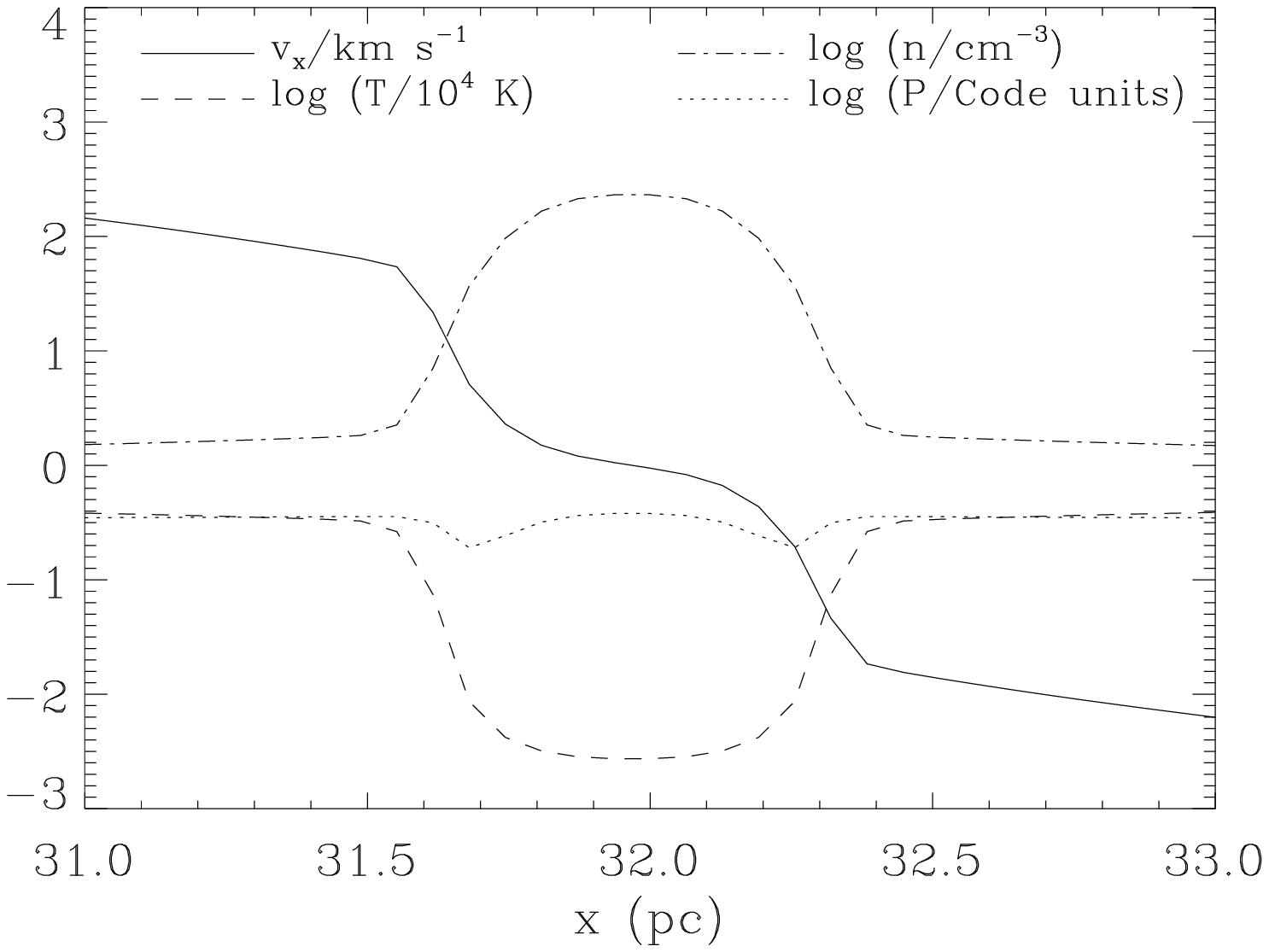}{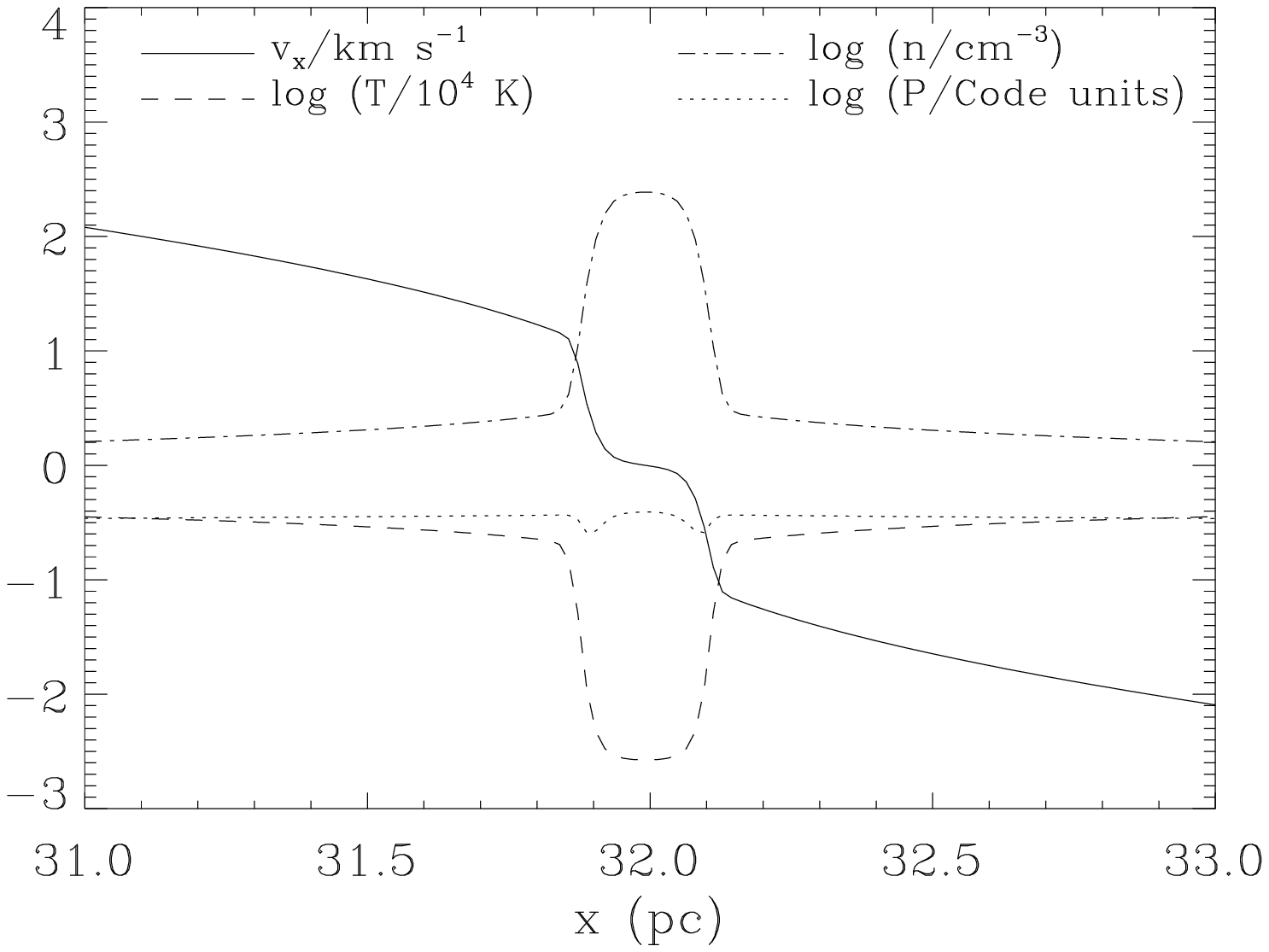}
\caption{Same as fig.\ \ref{fig:Mach0.96} but for run M1.03L64-1D at
$t=16.0$ Myr (\emph{left}) and run M1.03L64-1Dhr at $t=8.0$ Myr
(\emph{right}). Each run is seen to reach the converged
value of the density ($\sim 240 \pcc$) at a different time.} 
\label{fig:Mach0.96_interm}
\end{figure}

The fact that the cold layer thickness is initially resolution-limited also
implies that the density within it is artificially reduced, since the
layer is thicker than it should, and the total mass in the layer is
determined only by the accretion rate, which is independent of the
resolution. In turn, this implies that the pressure is below 
its physical value, as clearly seen in fig.\ \ref{fig:Mach0.96}. The
pressure in the cold layer can only reach the equilibrium value with its
surroundings at later times when the slab thickness is enough for
the density to reach its physical value. This happens at times $\sim$
16.0 and 8.0 Myr for runs M1.03L64-1D and M1.03L64-1D-hr,
respectively, at which the slab can be considered to start being well
resolved. This condition is shown in fig.\
\ref{fig:Mach0.96_interm}, in which the peak density is seen to have
converged in the two runs at different times.

However, since the total mass in the layer is independent of the resolution,
we expect a similar situation for the layer's column density.
Indeed, the thickness of the slab in run M1.03L64-1D at $t=5.33$
Myr (fig.\ \ref{fig:Mach0.96}) is $\sim 9$ grid zones, or
$\sim 0.58$ pc, with a maximum density of $\sim 99.7$ cm$^{-3}$. The
column density through the layer is measured at $N \approx 7.6 \times
10^{19}$ cm$^{-2}$. 
This can be compared with the estimates of \S
\ref{sec:early}. For $\Mr =1$, the model gives $\vf = 0.014$ km s$^{-1}$
and $n_3 = 240$ cm$^{-3}$, so that 
after $\Delta t = 5.33$ Myr, the layer thickness should be $\ell = 2 \vf
\Delta t\approx 0.15$
pc, and the column density $N \approx 1.125 \times 10^{20}
\psc$. Thus, the column density in the simulation is within a $\sim
30$\% error of the predicted value, even though the layer is $\sim 3.9$
times thicker than predicted at this time. For run M1.03L64-1D-hr,
the slab thickness is $\sim 15$ zones, or $\sim 0.24$ 
pc, and the maximum density is $218$ cm$^{-3}$, so the layer is close to
being well resolved, but not quite yet. The measured column density is
$N \approx 6.8 \times 10^{19}$, within a $\sim 40$\% error of the
predicted value. 
The column densities measured in the simulations at this early time are
lower than the predicted value at this time because the layers in the
simulations have actually bell-shaped profiles, rather than the flat ones
that develop at later times, at which the agreement with the model
predictions is much better (see \S \ref{sec:mod_sim_comp} and fig.\
\ref{fig:M0.96_mod_sim_compar}).

From the above discussion we conclude that the resolution of the
simulations in this paper is generally insufficient for correctly
resolving the structure inside the cold layer during the initial stages
of the evolution. Nevertheless, the column density of
the cold layer is reasonably 
recovered even from the early stages of the simulations.

\begin{figure}
\epsscale{1.}
\plottwo{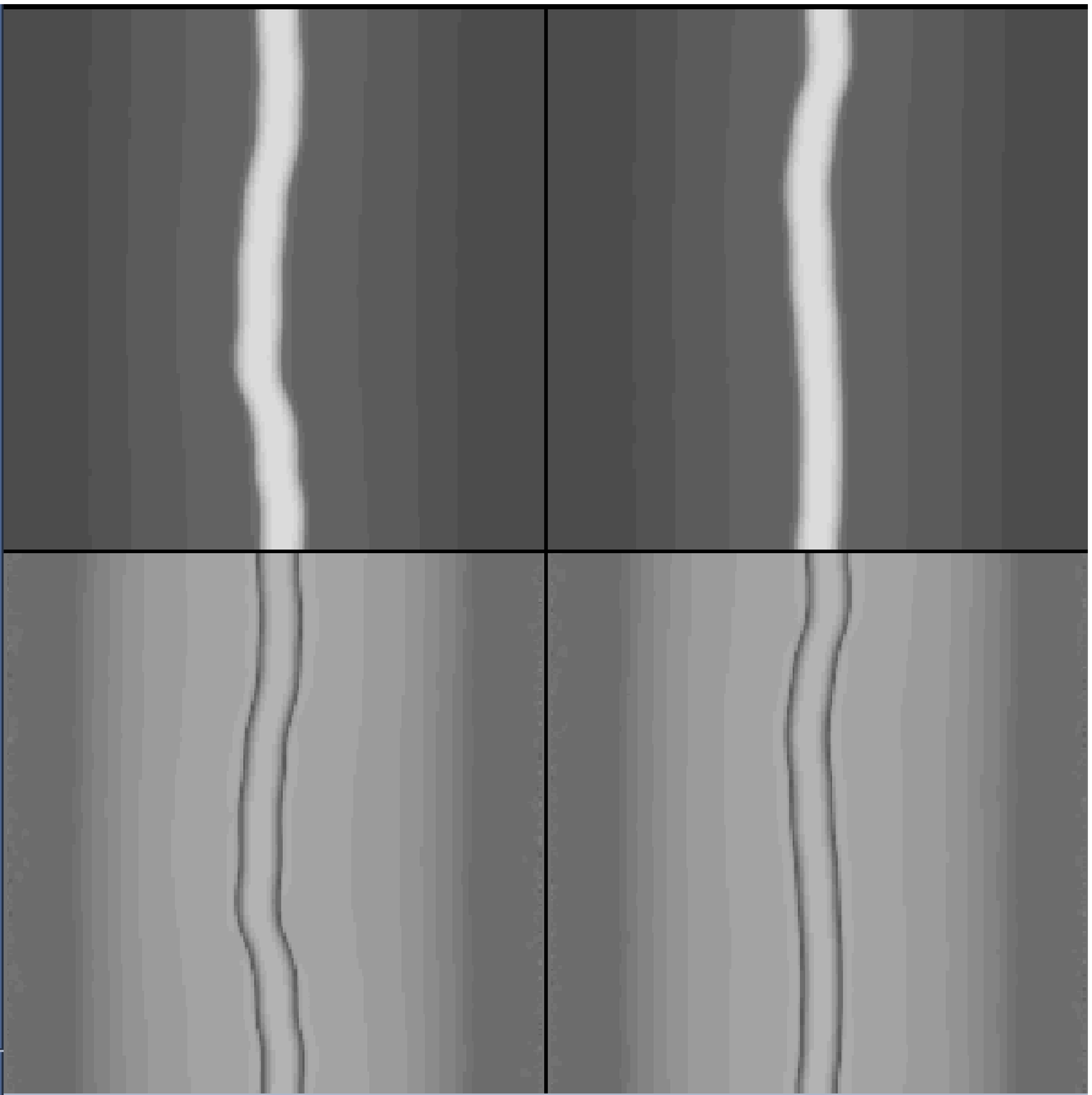}{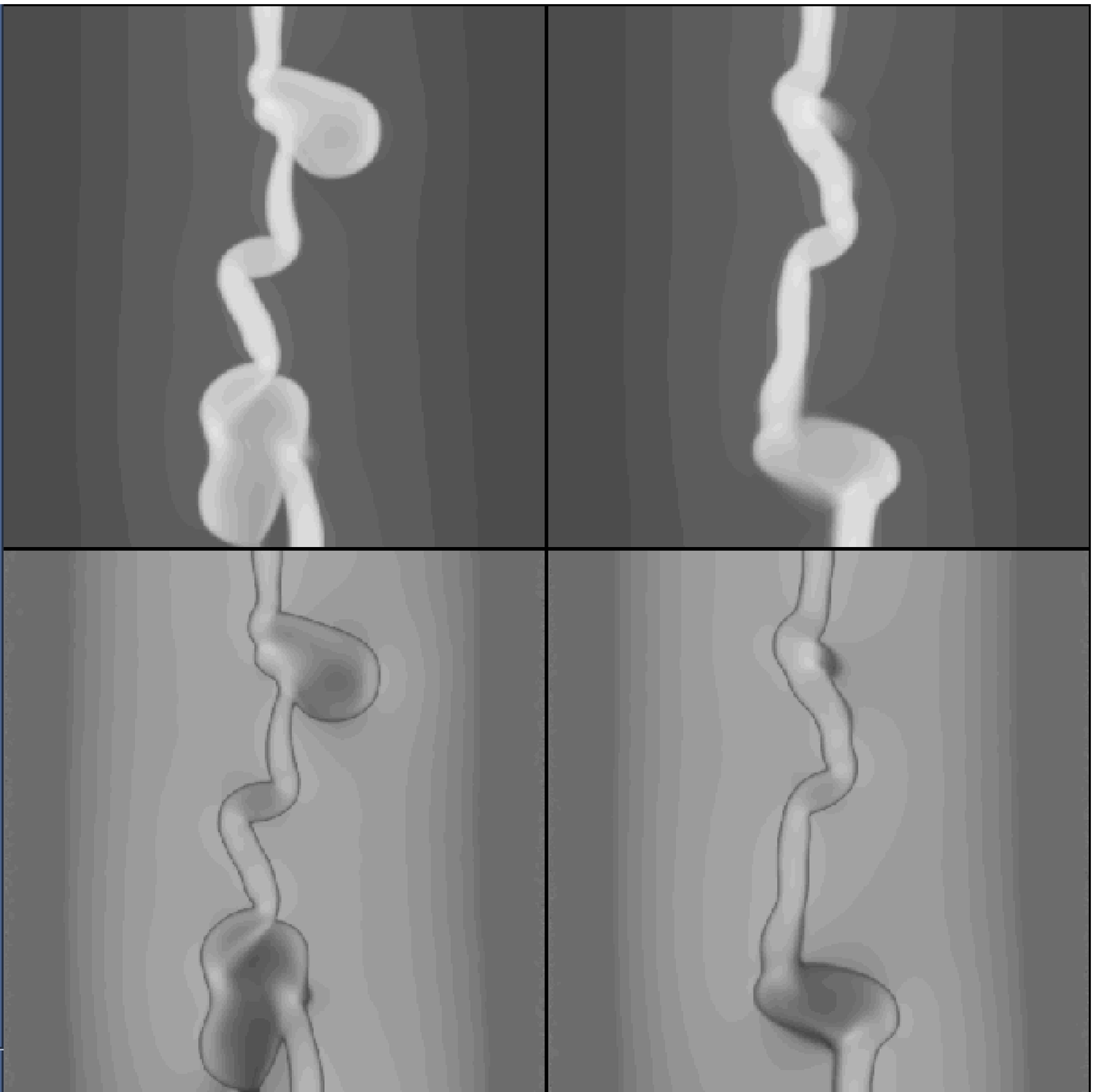}
\caption{Cross-section views of the density (\emph{upper rows}) and the
pressure (\emph{lower rows}) at $z=25$ (\emph{left columns}) and at
$z=50$ (\emph{right columns}) of runs M1.03L64 (\emph{left}) and
M1.03L64hr (\emph{right}) at $t=106.6$ Myr. The density and pressure
ranges are $(\rho_{\rm min},\rho_{\rm max}) = (0.34,223)\pcc$, $(P_{\rm
min},P_{\rm max}) = (1800,6200)$ K $\pcc$ for run M1.03L64 and
$(\rho_{\rm min},\rho_{\rm max}) = (0.34,370)\pcc$, $(P_{\rm min},P_{\rm
max}) = (1190,8100)$ K $\pcc$ for run M1.03L64hr. 
The higher-resolution run has already started to develop
turbulence, while the lower-resolution one is only undergoing slab
bending at this time.)} 
\label{fig:Mach0.96_fin}
\end{figure}

With respect to the late stages of the evolution, we have found
empirically that as the resolution is increased, the instability in the
compressed layer develops earlier. This is
shown in fig.\ \ref{fig:Mach0.96_fin}, which depicts constant-$z$ slices
of the density (\emph{upper panels}) and of the pressure (\emph{lower
panels}) for runs M1.03L64-LZ (\emph{left}) and M1.03L64-hr-LZ (\emph{right})
at the last computed step ($t=106$ Myr). It is seen that the instability
has already developed in the high-resolution run, but not quite yet in
the low-resolution one, although in this run the cold layer is already
significantly bent, and presumably turbulence will be generated 
soon after this time. Since our 3D simulations are still far from resolving
the true Field length in the various ISM regimes, the effect of
numerical diffusion roughly approaches that of the correct heat conduction as
the resolution is increased. Therefore, the observed trend of decreasing
turbulence development time with increasing resolution implies that the
timescales reported in this paper should be considered upper limits
only.

Concerning the structures formed by the turbulence in the compressed
layer, once it has developed fully, the main limitation introduced by the
limited resolution is that the size of the cold structures will be
artificially bounded from below by the numerical cell size, but we
expect no other serious limitation. The structures are mainly formed by
the turbulent flow, not by TI, and therefore they can in principle form with
a variety of sizes, rather than at the characteristic scales of the TI,
which indeed are very small.


Finally, we note that, as seen from figs.\ \ref{fig:M1.2_movie}
and \ref{fig:M2.4_movie}, in runs M1.2L32 and M2.4L16, as the
turbulent layer thickens, its bounding shock eventually touches the
inflow boundary. This happens at times $t=42.6$ Myr for run M1.2L32
(frame 32 in the corresponding animation, fig.\ \ref{fig:M1.2_movie},
with the frame-count starting at frame zero), and $t=7.34$ Myr for run
M2.4L16 (frame 11 in fig.\ \ref{fig:M2.4_movie}). 
At this point, the simulations are in principle not valid
anymore, as the interaction of the material in the turbulent layer with
the inflowing gas ceases to be followed in full. In practice, however, we have
found that the statistical properties of the simulation are not affected
by the collision of the shock with the boundary. This fact can be
understood because the inflow boundary conditions effectively act as
outflow conditions for the material reaching the boundary from the
inside of the box, and so effectively this gas just leaves the box as
through standard outflow conditions, while fresh gas continues to flow
in through these boundaries, and to interact with the gas remaining in the
simulation. Thus, in \S \ref{sec:turb_gas} we discuss various physical
and statistical properties both at the time the shocks leave the
simulations, and at the time when the statistics become stationary.

\subsection{Results} \label{sec:results}

\subsubsection{Comparison with the analytical model}
\label{sec:mod_sim_comp}

The predictions of the analytical model of \S \ref{sec:early} can
be compared with the results of the 1D numerical simulations, which
resolve the layer at not-too-late times, and in which the slab does not
become unstable. 
To be consistent with the hypotheses of the model, we consider 
times late enough that the outer shock has essentially stopped.

\begin{figure}
\epsscale{1.}
\plottwo{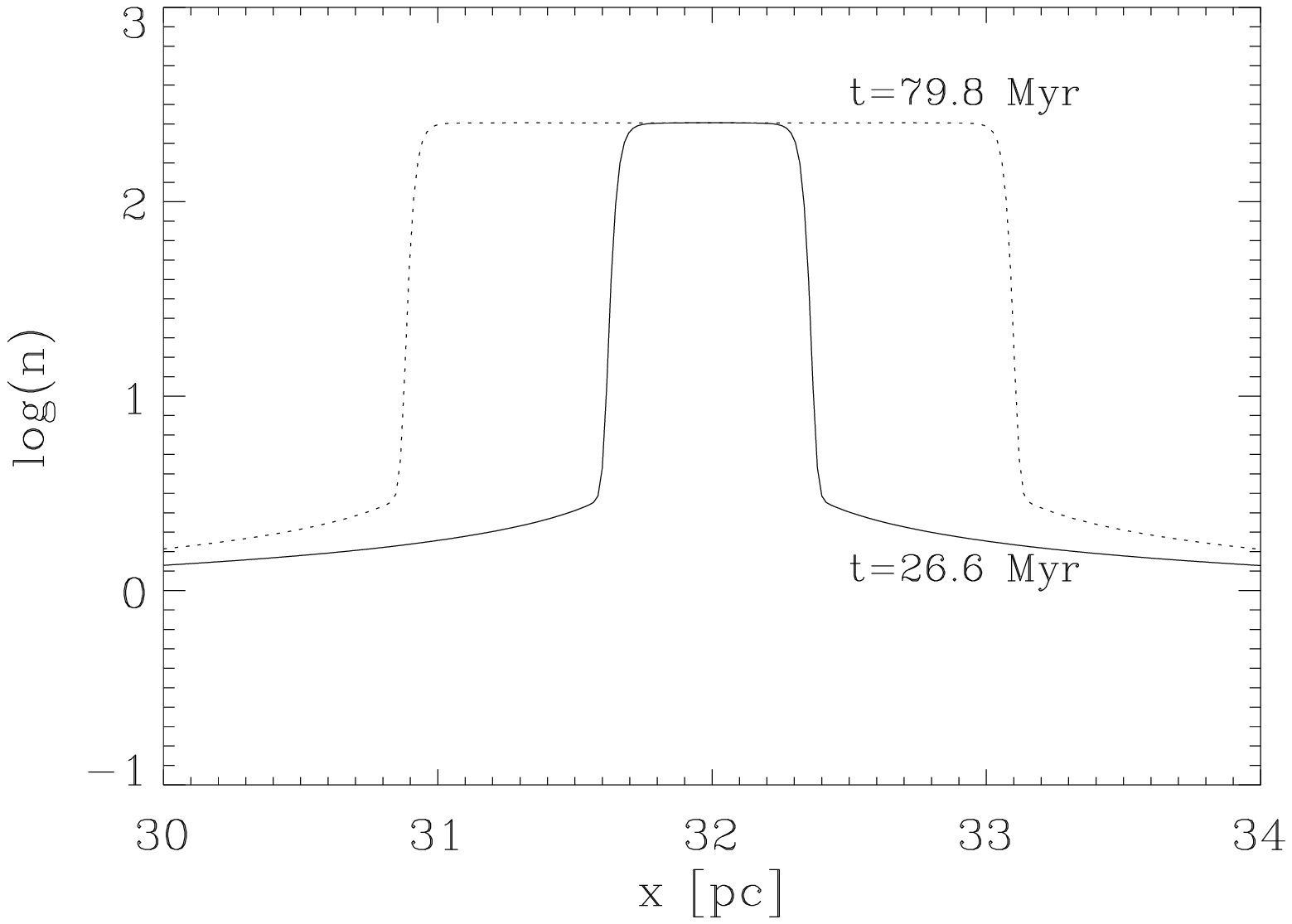}{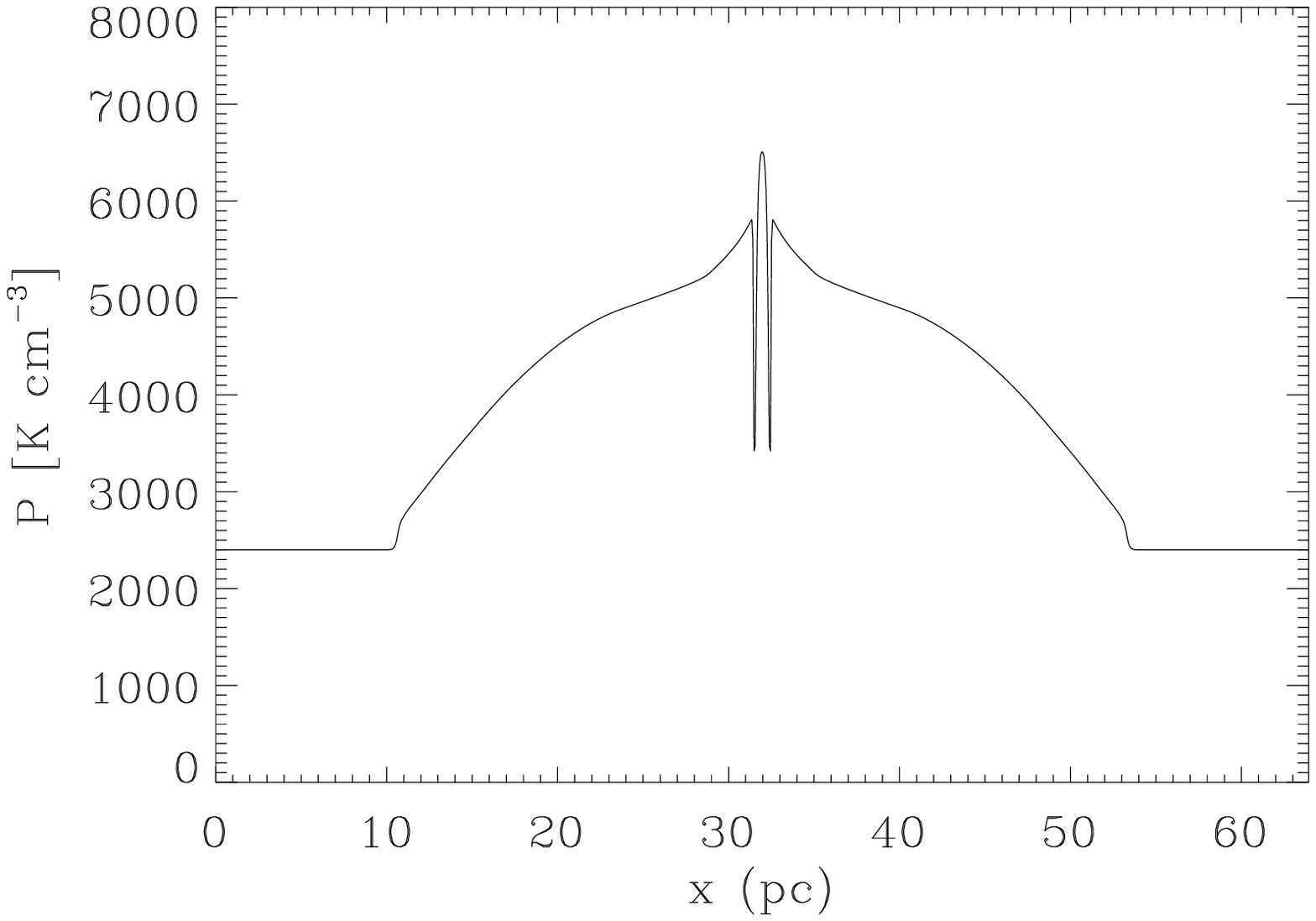}
\caption{\emph{Left:} Number density profile in the centermost region of
run M1.03L64-1Dhr at times 26.6 Myr (\emph{solid line}) and 79.8 Myr
(\emph{dashed line}), allowing measurement of the front expansion
velocity $\vf$. \emph{Right:} Pressure profile of the same run
over the entire length of the simulation at $t=26.6$ Myr.} 
\label{fig:M0.96_mod_sim_compar}
\end{figure}

As illustrations, we discuss the cases with $\Mr = 1.03$ and
1.2. The left panel of fig.\ \ref{fig:M0.96_mod_sim_compar} shows the
density field in the central 6 pc of run M1.03L64-1Dhr at times 26.6 Myr
and 79.8 Myr. We can see that the right-hand side of the dense cold
layer has moved 0.8 pc, from $x=32.3$ pc to $x=33.1$ pc. This gives a
velocity $\vf = 0.015 \kms$. We also read the density inside the layer
as 255$\pcc$. The right panel of fig.\ \ref{fig:M0.96_mod_sim_compar}
shows the pressure throughout the entire simulation at $t=26.6$
Myr. The maximum value of the pressure, occurring at the center of the
dense layer, is $P_3 = 6653$ K$\pcc$. This can be compared with 
the model predictions for $\Mr =1.03$, which are $P_3 = 6643$ K$\pcc$,
$n_3 = 257 \pcc$ and $\vf = 0.014 \kms$, in 
excellent agreement with the measurements.

\begin{figure}
\epsscale{1.}
\plottwo{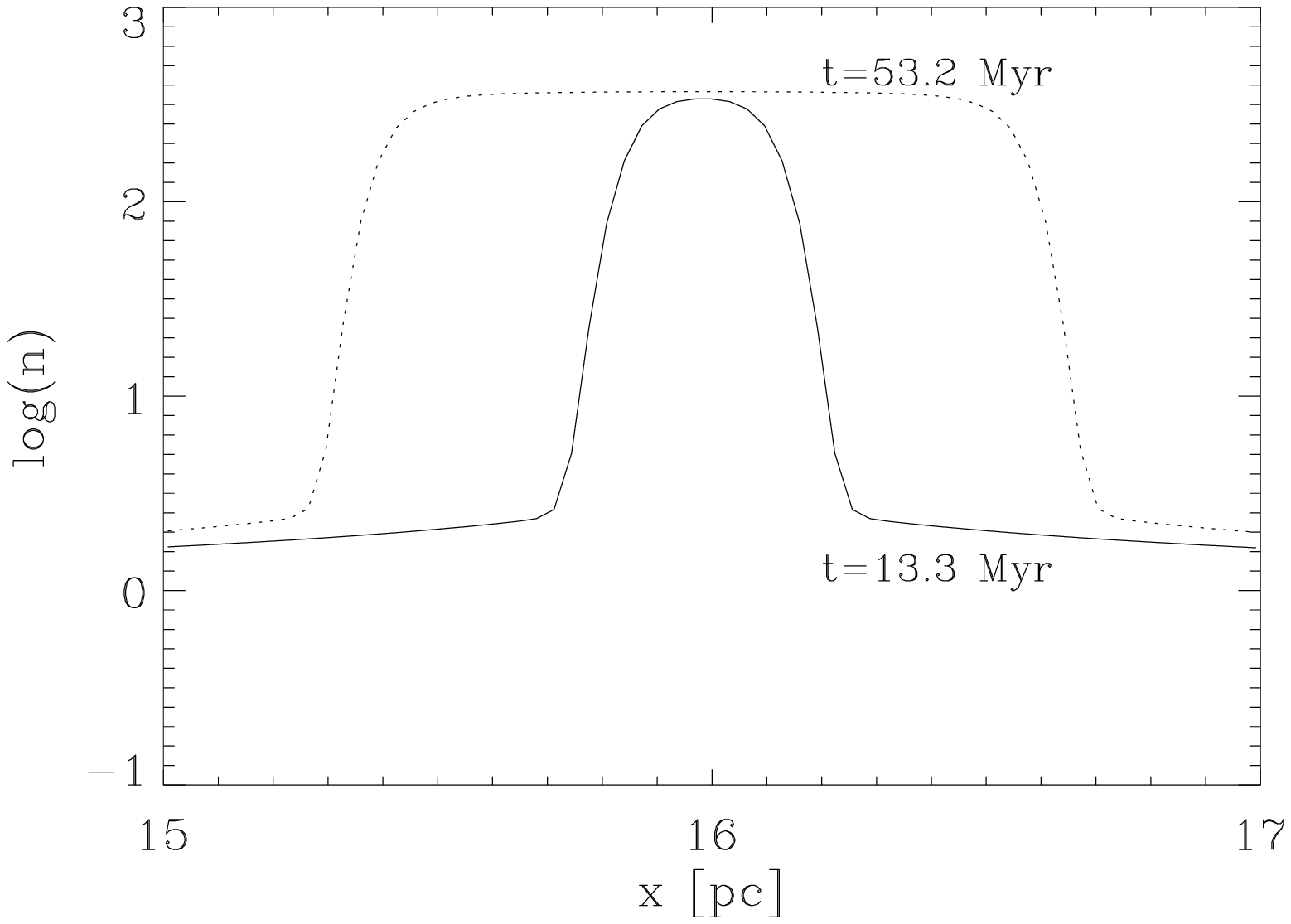}{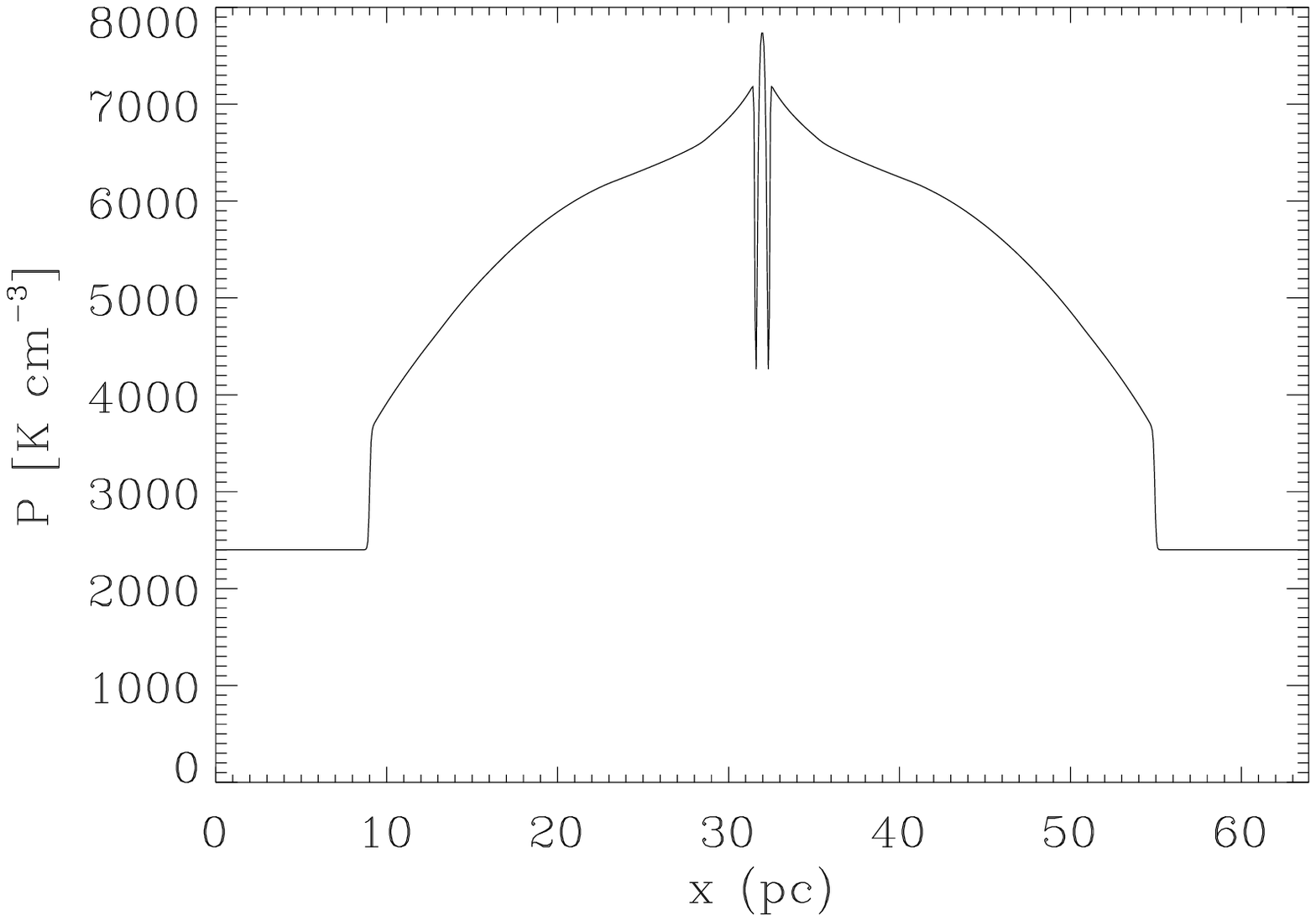}
\caption{Same as fig.\ \ref{fig:M0.96_mod_sim_compar} but for run
M1.2L321D. The right frame shows the pressure field at $t=13.3$ Myr.} 
\label{fig:M1.2_mod_sim_compar}
\end{figure}

Figure \ref{fig:M1.2_mod_sim_compar} shows the corresponding
comparison in the case $\Mr=1.2$. In this case, the left panel shows
that the front moves from $x = 16.20$ pc at $t= 13.3$ Myr to $x = 16.65$
pc at $t=53.2$ Myr,
implying $\vf= 0.011 \kms$. We also read a cold layer density of $n_3=
338 \pcc$. In turn, the right panel shows the pressure, whose maximum
value is $P_3 = 7734$. For a Mach number $\Mr = 1.2$, our model
gives $P_3 =8160$ K$\pcc$, $\vf =  0.0106 \kms$ and $n_3 = 378
\pcc$ (cf.\ fig.\ \ref{fig:ncold_Minfl}). This is again in good
agreement with the results of the simulation.

We conclude that the analytical model and the 1D numerical
simulations are consistent with each other, giving us confidence in
both,
and confirming the physical scenario (an
outer shock front and an inner condensation front), in which
thin sheets form during the initial stages of the compression.

\subsubsection{Global evolution} \label{sec:global_evol}

\begin{figure}
\epsscale{1.}
\plottwo{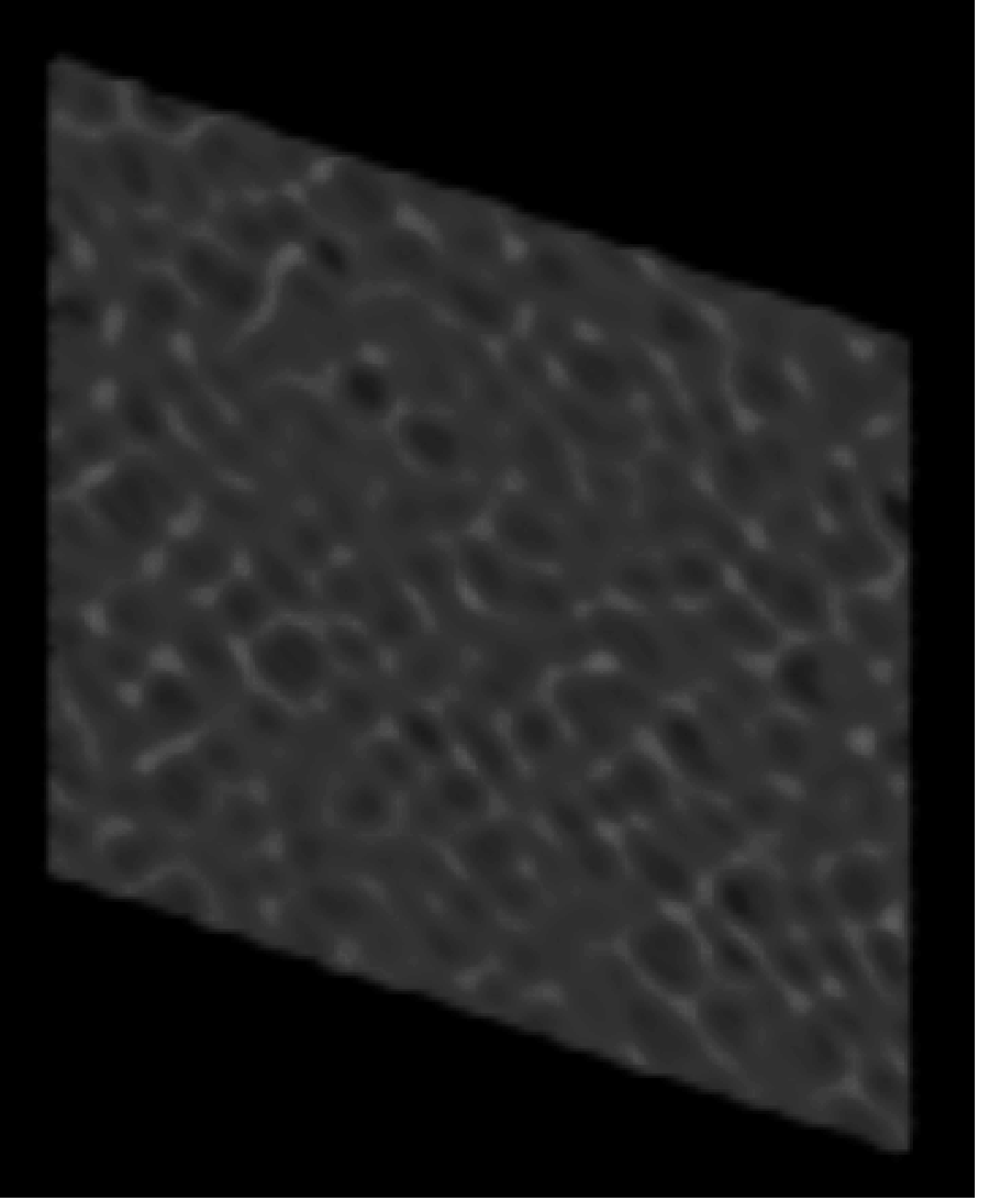}{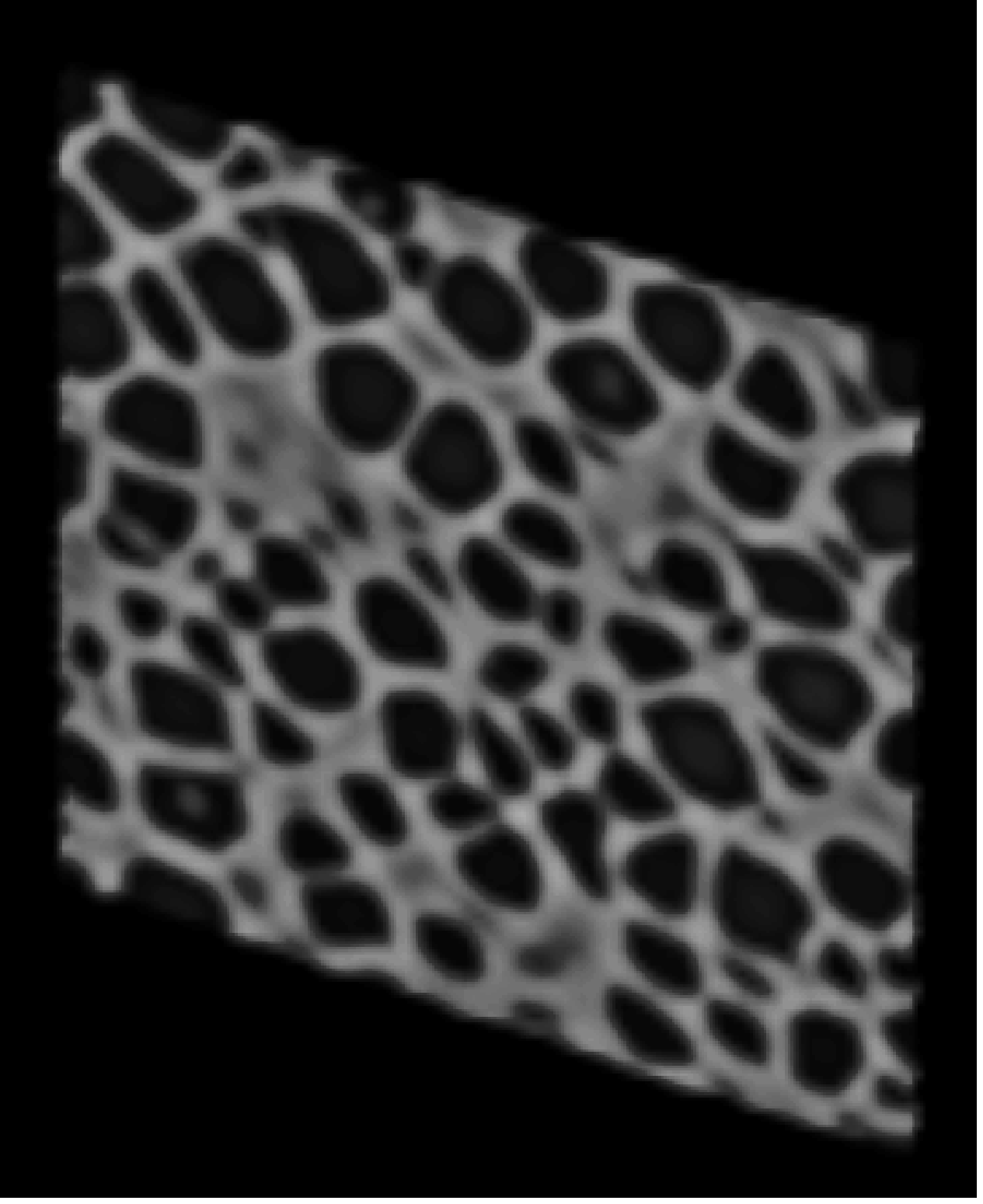}
\caption{Projection views of the density field in run M1.03L64 at
$t=5.33$ Myr (\emph{left}, corresponding to frame 2 in the animation of
the entire 
evolution available in the electronic edition of the \emph{Astrophysical
Journal}, with the frame-count starting at zero) and at $t=10.67$ Myr
(\emph{right}, frame 4). The thin sheet is seen to fragment into a
honeycomb pattern. }
\label{fig:M0.96_movie}
\end{figure}

\begin{figure}
\plotone{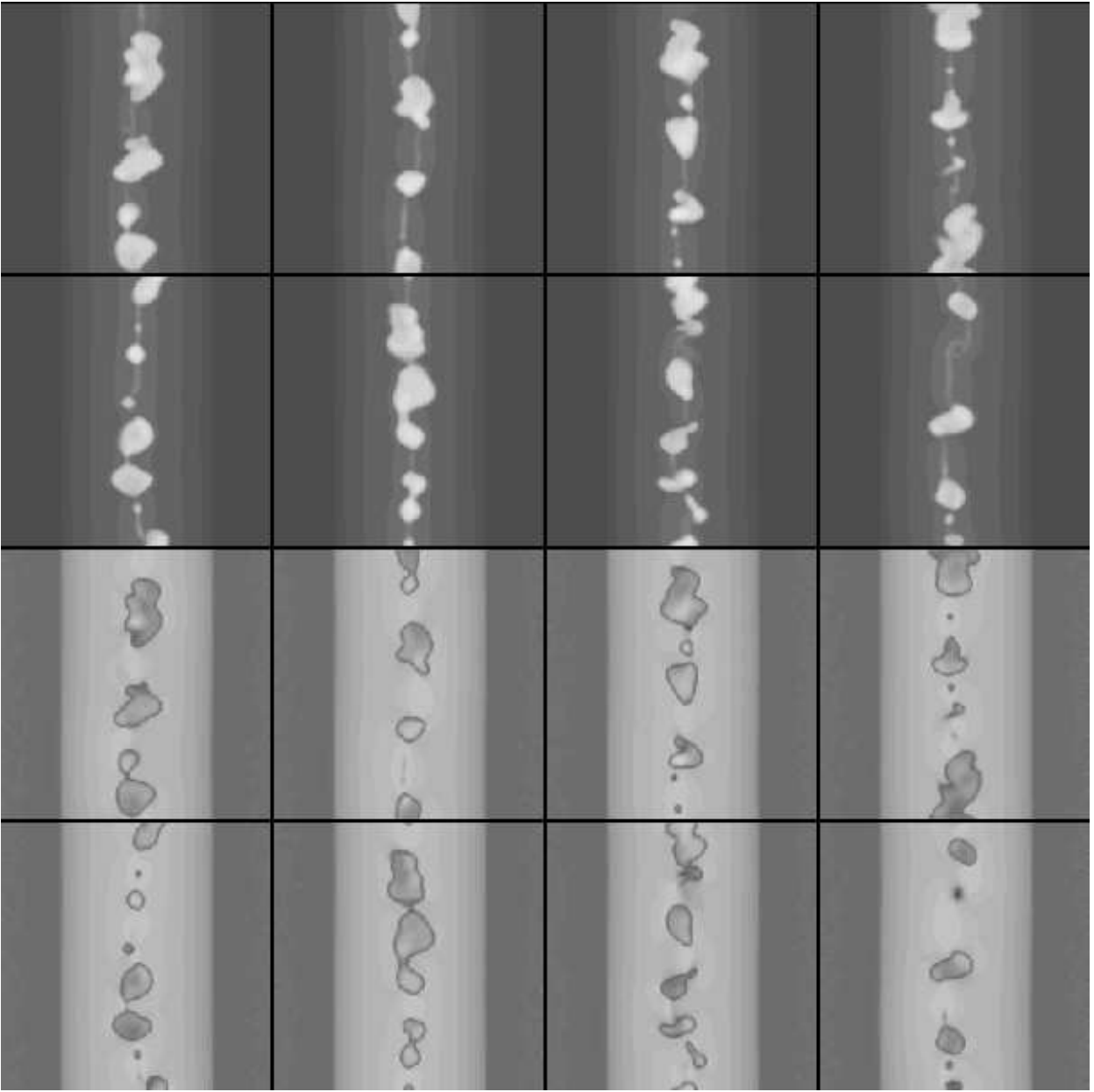}
\caption{Cross-section views at eight $z$ values, spaced by $\Delta z =
25$ grid cells, of the density (\emph{upper two rows}) and pressure
(\emph{lower two rows}) in run M1.2L32 at $t=20$ Myr. The dynamic
ranges of the density and pressure are indicated by the \emph{left} 
panels of figs.\ \ref{fig:rho_hists} and \ref{fig:P_hists}, respectively.
The still image in the printed version 
corresponds to frame 15 in the animation available in the
electronic edition of the \emph{Astrophysical Journal}. 
In the animation, the density and pressure frames are reversed.
} 
\label{fig:M1.2_movie}
\end{figure}

\begin{figure}
\plotone{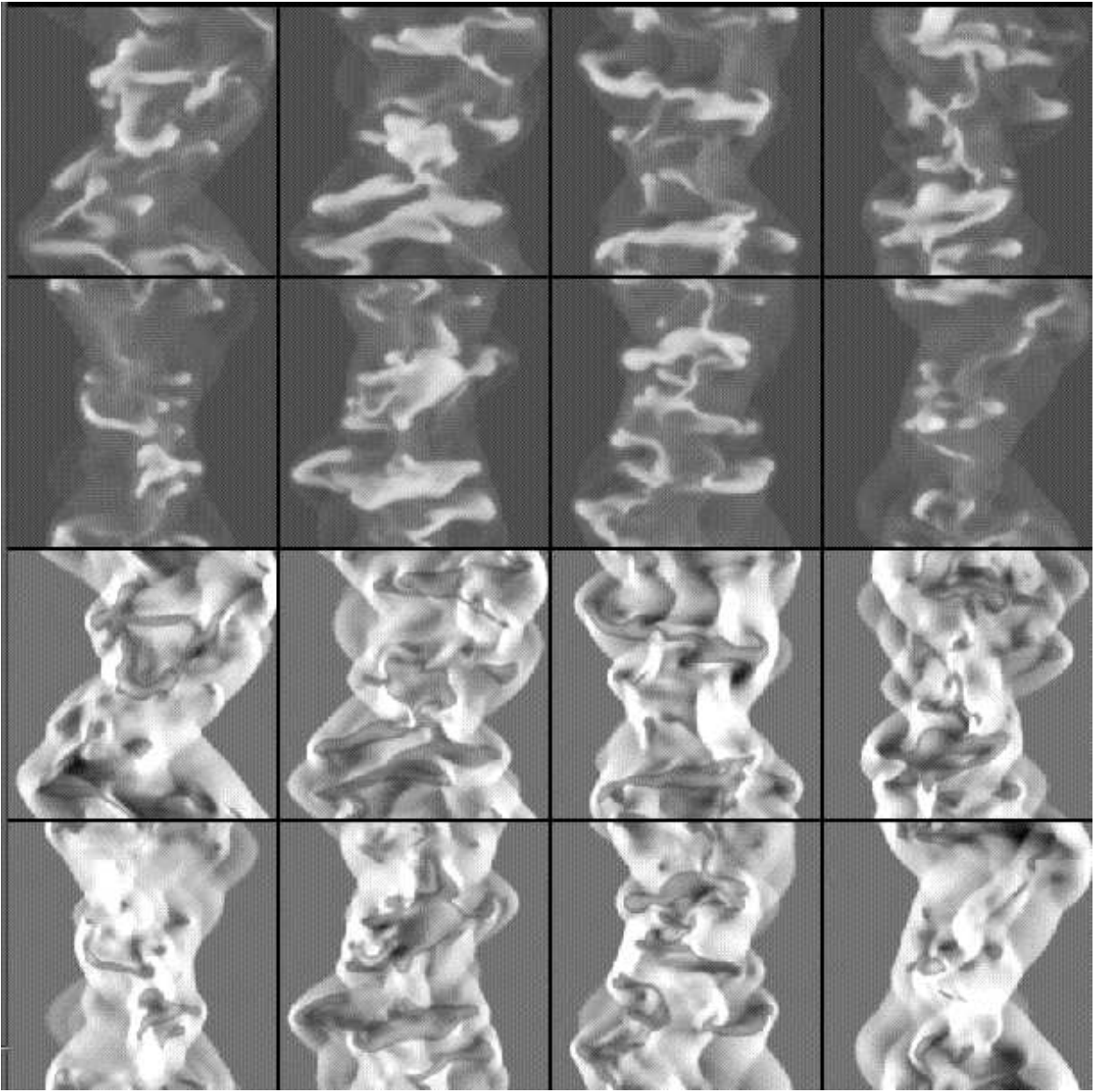}
\caption{Same as fig.\ \ref{fig:M1.2_movie} but for run M2.4L16. The
dynamic ranges of the density and pressure are indicated by the \emph{right} 
panels of figs.\ \ref{fig:rho_hists} and \ref{fig:P_hists}, respectively.} 
\label{fig:M2.4_movie}
\end{figure}

In this section we now discuss the evolution of the compressed and the
cold layers in runs with various inflow Mach numbers. In the animations of
figures \ref{fig:M0.96_movie}, \ref{fig:M1.2_movie} and \ref{fig:M2.4_movie} 
we respectively show the evolution of runs
M1.03L64, M1.2L32 and M2.4L16. The spacing between
frames in the animations is $\Delta t = 0.5$ code time units (cf.\ Table
\ref{tab:runs} for the time unit in each run), which corresponds to
$\Delta t =$ 2.66, 1.33 and 0.66 Myr, respectively. The animations in
figures \ref{fig:M1.2_movie} and \ref{fig:M2.4_movie} show constant-$z$
cross-sections of the simulation at different, equally-spaced $z$
values, with $\Delta z = 25$ grid cells, to illustrate the evolution of
the thickness and structure of the dense ``layer'' (which, in the
late stages, becomes a turbulent cloud). The animation in fig.\
\ref{fig:M0.96_movie} shows instead a transluscent projection of the
density evolution in run M1.03L64, to illustrate the fragmentation
process of the thin dense layer. The still figures of the printed
version show selected panels from the animations 
to also illustrate these features. 
%
%

From the animations and figures, several points are noticed. In general,
we see that, after the formation of the thin cold layer, the latter
begins to fragment into a filamentary honeycomb pattern, with
denser clumps at the sites where the filaments intersect. Subsequently,
the density structures begin to move on the plane of the dense layer,
merging and forming larger clumps. However, apparently the random
fluctuations in the inflow velocity cause the merging clumps to collide
with slight offsets, which therefore cause the layer to thicken and to
develop vorticity. Ultimately, the motion in the cold thin
layer appears to completely destabilize the entire thick shocked slab,
and fully developed turbulence ensues. 

In fig.\ \ref{fig:M1.2_movie} it is interesting to note that the
density peaks (``clouds'') appear surrounded by a low-pressure interface
in the pressure images. This region probably 
corresponds in our simulations to a numerical effect, with
the pressure gradient being compensated there by numerical diffusion,
although in real clouds this may correspond to a conducting 
interface. This interface, however, is not apparent in the pressure
images for run M2.4L16 (fig.\ \ref{fig:M2.4_movie}). These results
suggest that sharp phase transitions between the warm and cold gas still
exist  at $\Mr = 1.2$ \citep{AH05}, but tend to be erased at $\Mr =
2.4$, as also suggested by the pressure histograms discussed in \S
\ref{sec:rho_P_dists}. Nevertheless, even in run M1.2L32, the
pressure in the ``clouds'' is seen to fluctuate significantly, because
of their dynamical origin, and in general they are not all at the same
pressure nor at uniform pressure inside. 

An important datum of the simulations is the time they require for
attaining a saturated turbulent state. This can be defined in
practice as the timescale for reaching a stationary shape of the
statistical indicators, such as the density and pressure histograms. We
find that this occurs at $t \approx 47.9$ Myr for run M1.2L32 (frame 36
in the corresponding animation, fig.\ \ref{fig:M1.2_movie}), and at $t
\approx 10.7$ Myr for run M2.4L16 (frame 11 in fig.\
\ref{fig:M2.4_movie}). Run M1.03L64 does not seem to have reached a
stationary state by the end of the integration time we have considered,
$t=80.1$ Myr. As mentioned in \S \ref{sec:num_res}, these times are
larger than those
at which the shocks reach the boundary, and therefore in the next
section we discuss both.

\subsubsection{Properties of the turbulent state} \label{sec:turb_gas}

\subsubsubsection{Density and pressure distributions} \label{sec:rho_P_dists}

\begin{figure}
\epsscale{1.}
\plottwo{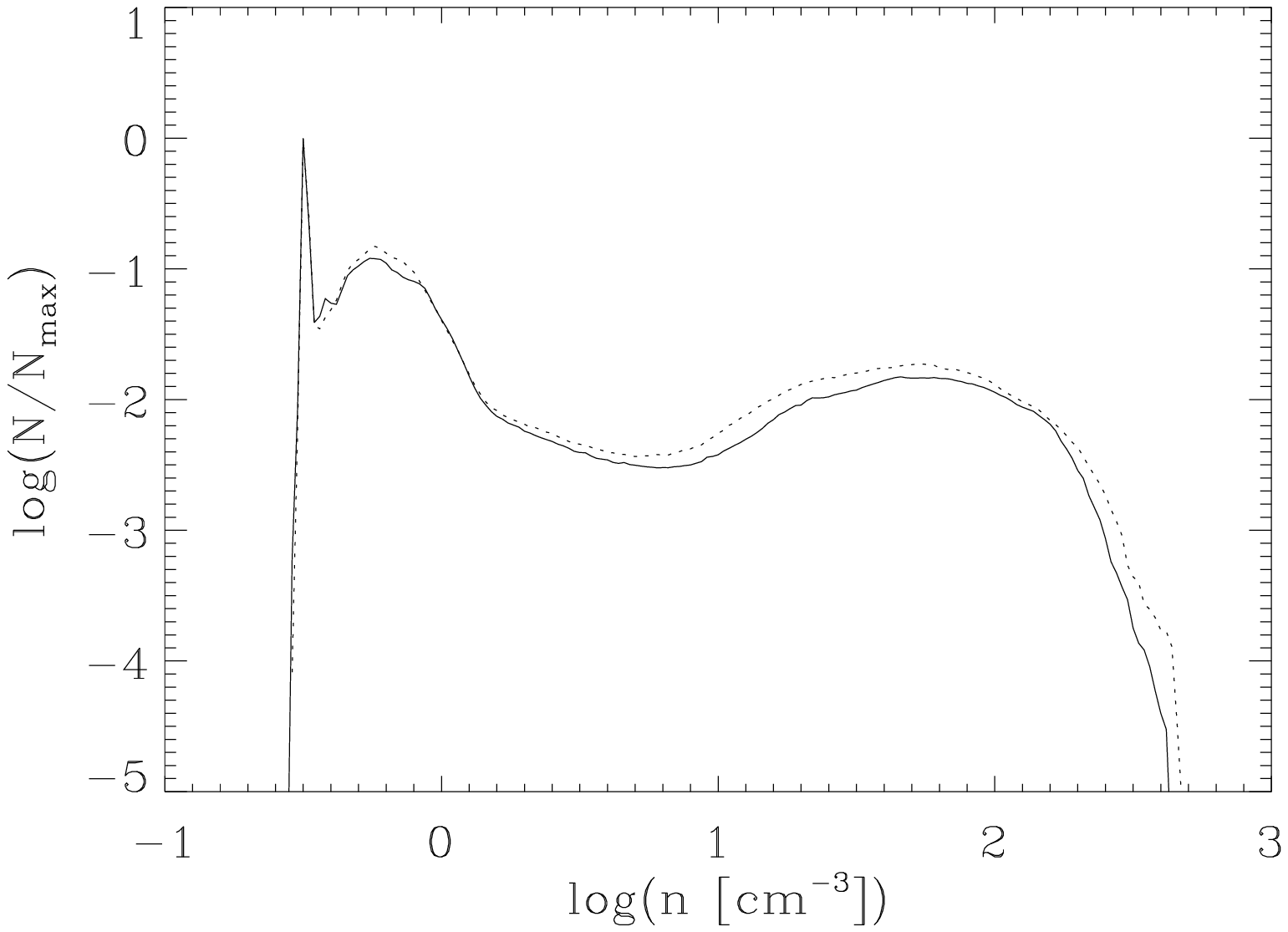}{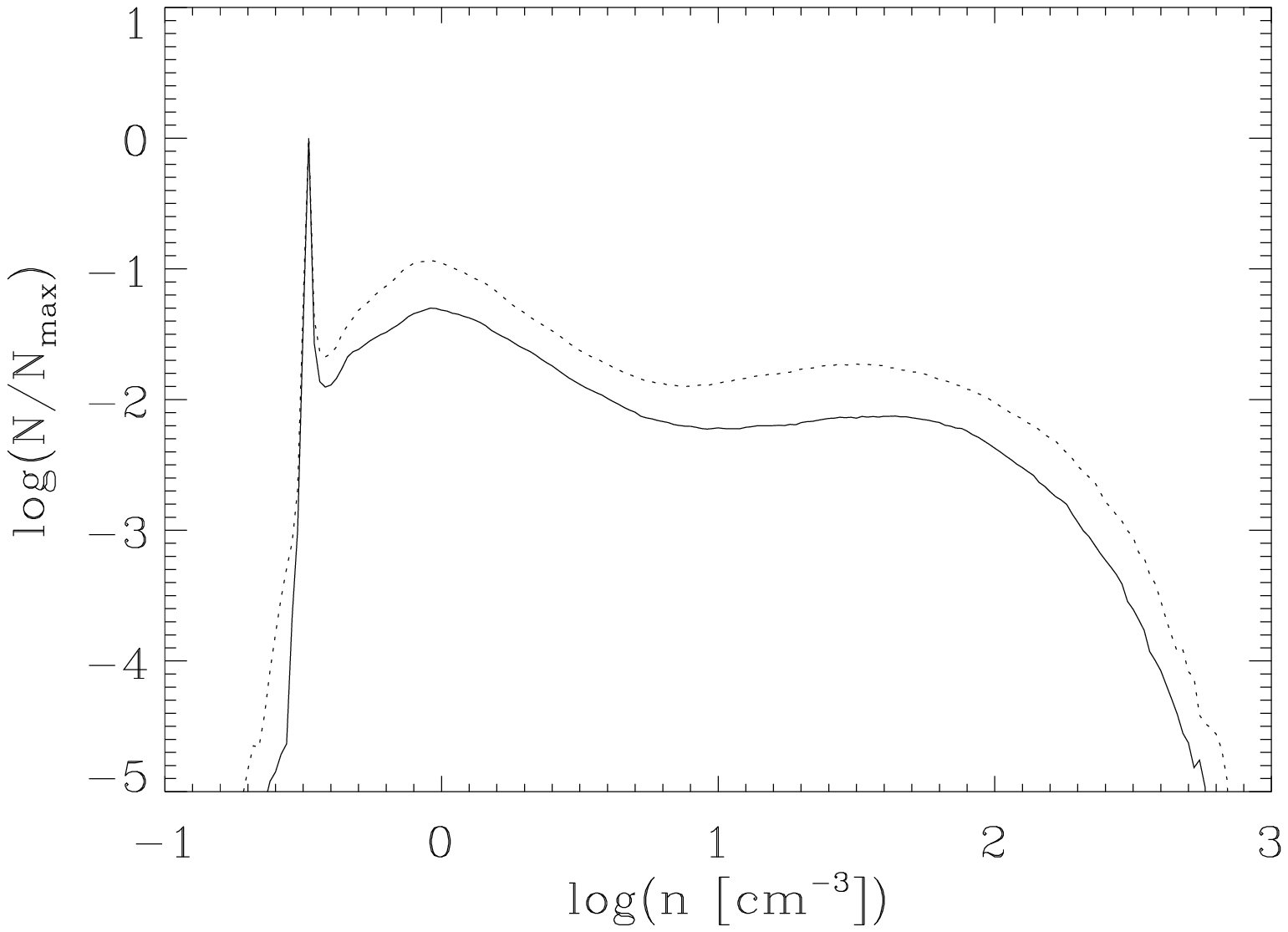}
\caption{\emph{Left:} Density histograms of run M1.2L32 at $t=42.6$
Myr (\emph{solid line}), at which the shock touches the $x$-boundaries,
and at $t= 47.9$ Myr (\emph{dotted line}). At the former time the
simulation is fully self-consistent but the turbulence is not completely
stationary yet. At the latter time, the reverse is true. The differences
between the two histograms are 
minimal, and consistent with the suggestion that the statistics are not
seriously affected by the shock leaving the simulation during the
turbulent stage. \emph{Right:}
Density histograms at the corresponding times for run M2.4L16, $t=7.37$
Myr (\emph{solid line}) and at $t= 10.7$ Myr, \emph{dotted line}).
}
\label{fig:rho_hists}
\end{figure}

In this section we concentrate on the cases with $\Mr=1.2$ and
$\Mr=2.4$, as they are the ones in which significant amounts of
turbulence can develop within realistic timescales in the shocked
layer. Figure \ref{fig:rho_hists}a shows the density histograms of run
M1.2L32 at $t= 42.6$ Myr (\emph{solid line}, frame 32 in the
animation of fig.\ \ref{fig:M1.2_movie}), when the shock touches the
boundary, and at $t= 47.9$ Myr, when the statistics become stationary
(\emph{dotted line}, frame 36 in the animation). The histograms at the
two times are very similar, although the former one contains slightly
lower numbers of grid cells with intermediate- and high-densities,
indicative of the not yet completely stationary turbulent regime at that
time. Similarly, fig.\ \ref{fig:rho_hists}b shows the density histograms
at the corresponding times for run M2.4L16 ($t= 7.37$ Myr, frame 11,
\emph{solid line}, and $t= 10.7$ Myr, frame 16, \emph{dotted line}). The
same trends are observed for this run.

The histograms of both runs have narrow and tall peaks at the density of
the unperturbed inflowing streams (the density of the warm phase,
$n_1=0.34~\pcc$). The histogram of the mildly supersonic run M1.2L32 is
significantly bimodal, although it extends to densities $\sim 400$
cm$^{-3}$, well into what is normally associated with typical molecular
cloud densities. The peak of the high-density maximum of the
distribution is between $\sim 50$ and $100~\pcc$.  In contrast, the
histogram for the strongly supersonic run M2.4L16 has a less pronounced
bimodal character, and extends at roughly constant height to densities
typical of the cold phase, to then start decreasing at higher densities,
to reach values close to 1000 cm$^{-3}$. Thus, a higher inflow Mach
number tends to erase the signature of bistability of the flow by
increasing its level of turbulence, in agreement with the studies by
\citet{SVG02}, \citet{AH05} and \citet{GVK05}. 

\begin{figure}
\epsscale{1.}
\plottwo{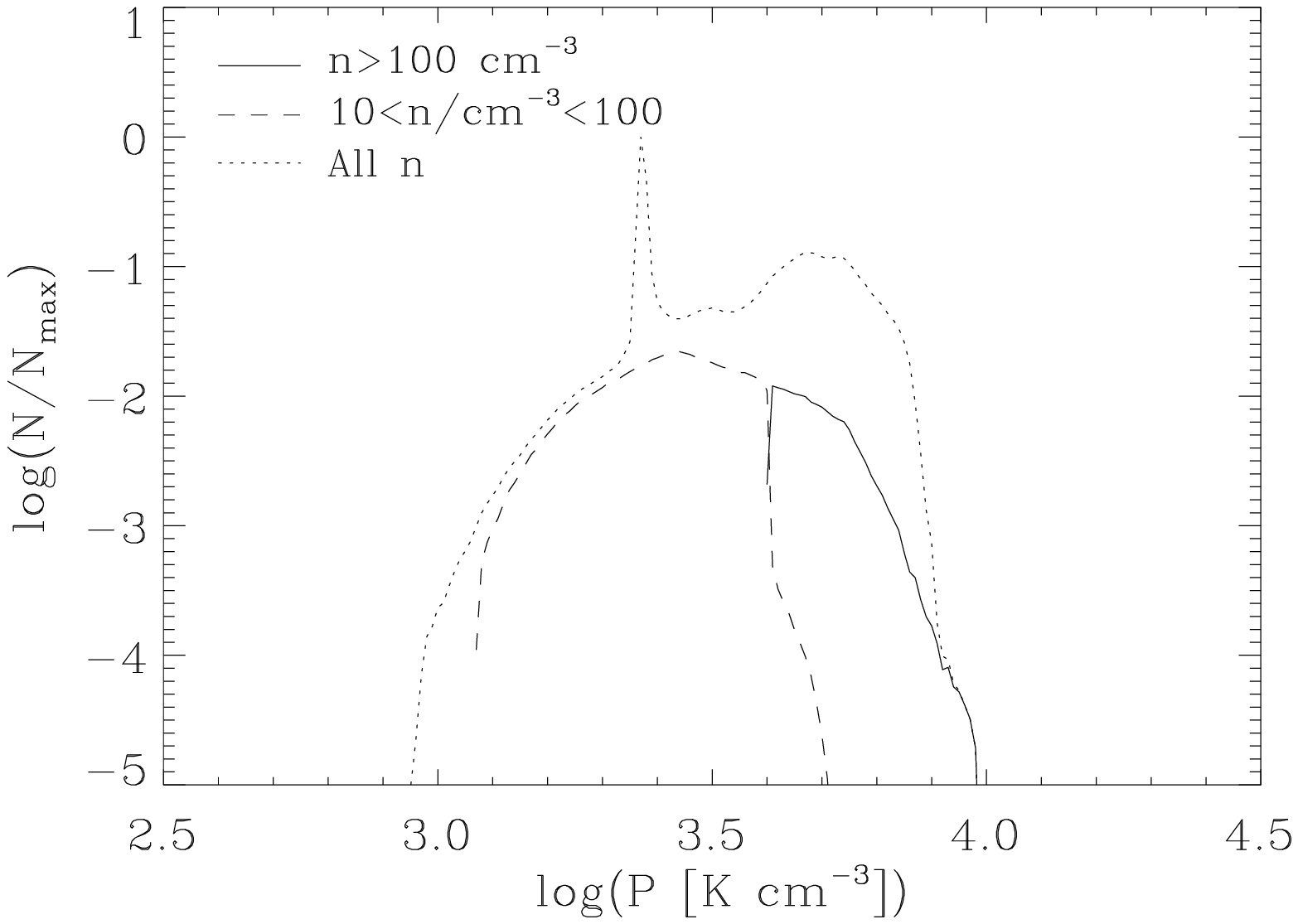}{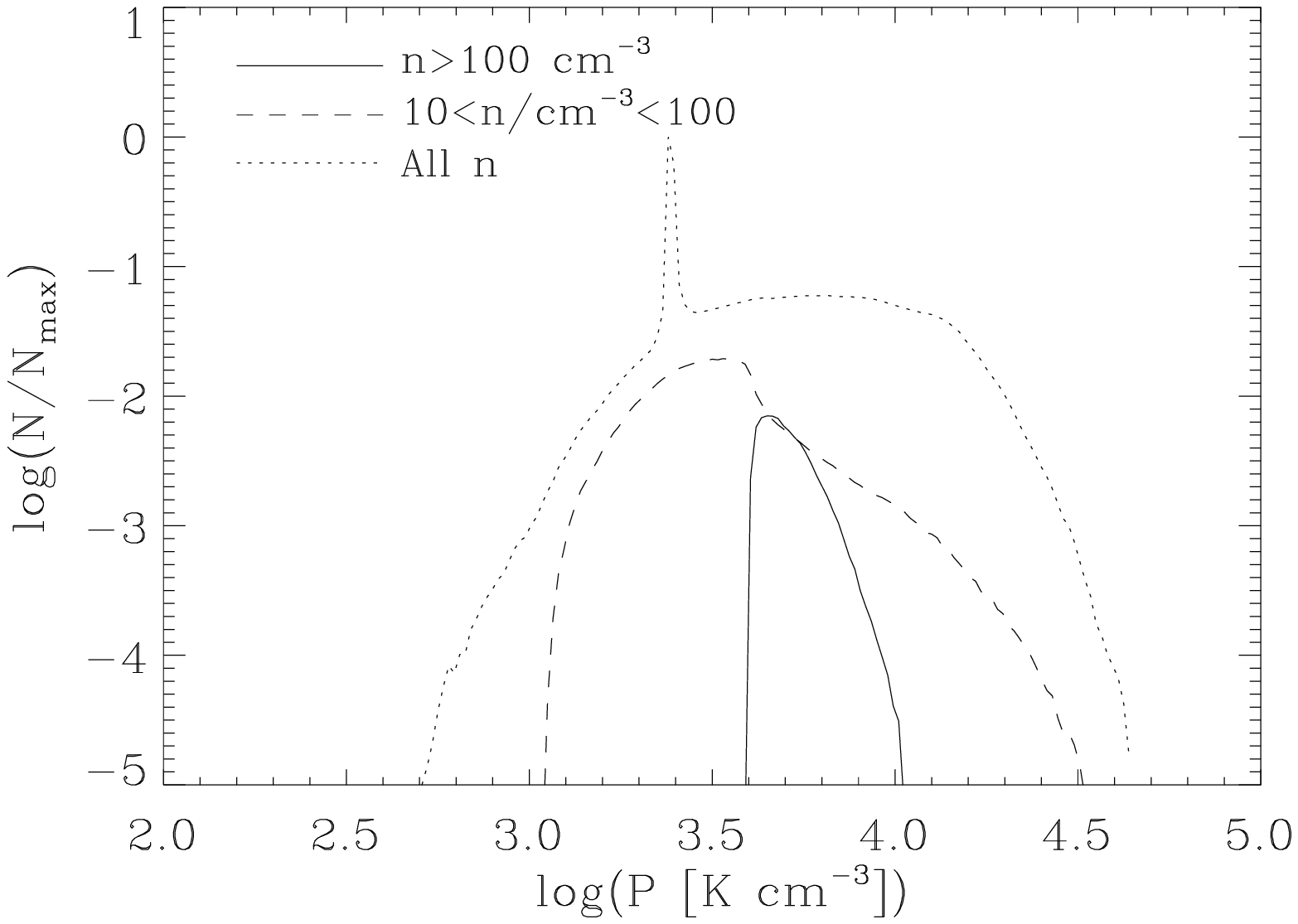}
\caption{Pressure histograms of runs M1.2L32 at $t=42.6$ Myr
(\emph{left}) and of run M2.4L16 at $t=7.37$ Myr (\emph{right}). The
\emph{dotted} line shows the total histogram, while the \emph{dashed}
line shows the histogram for the intermediate-density gas (IDG, with $10
\pcc < n < 100 \pcc$) and the \emph{solid} line shows the high-density
gas (HDG, with $n > 100 \pcc$).} 
\label{fig:P_hists}
\end{figure}

\begin{figure}
\epsscale{1.}
\plottwo{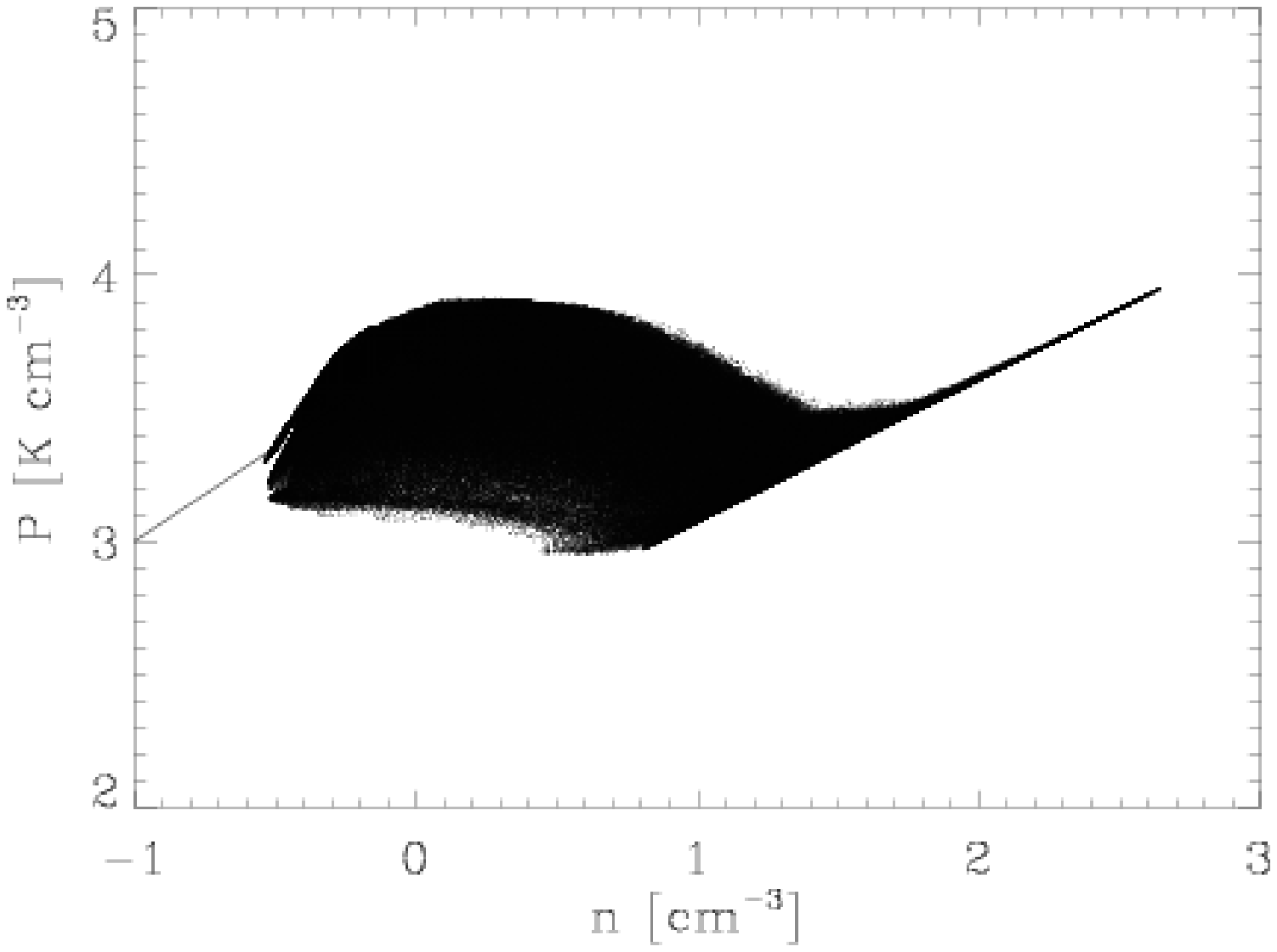}{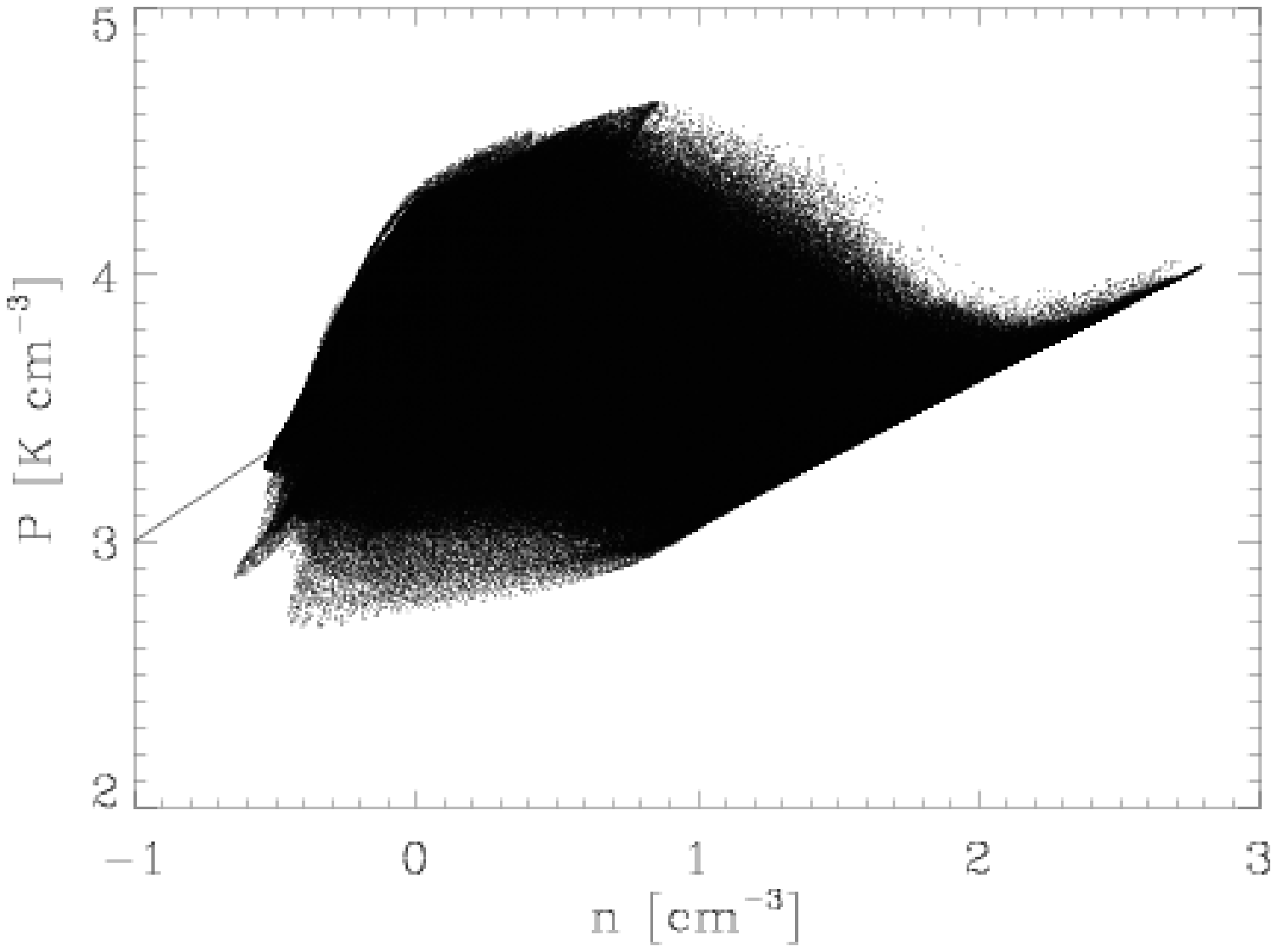}
\caption{Pressure-versus-density plots of runs M1.2L32 at $t=42.6$ Myr
(\emph{left}) and of run M2.4L16 at $t=7.37$ Myr (\emph{right}). The
\emph{solid} lines show the locus of the equilibrium pressure as a
function of density.} 
\label{fig:P_vs_rho}
\end{figure}

Of particular interest is the pressure distribution in these
simulations, as a means to understanding the overpressured nature of
molecular clouds. Figures \ref{fig:P_hists}a and b
show the pressure histograms of runs M1.2L32 (at $t=42.6$ Myr) and run
M2.4L16 (at $t=7.37$ Myr). Shown are the histograms for the entire
simulation (\emph{dotted} lines), for gas with densities 10  
cm$^{-3} < n < 100$ cm$^{-3}$ (\emph{dashed} lines), which we will
refer to as the ``intermediate-density'' gas (IDG), and for gas with
densities $n>100$ cm$^{-3}$ (\emph{solid} lines), which we will refer to
as ``high-density'' gas (HDG). The histograms for all components are
normalized to the total number of points. We will generally identify the
IDG with 
the CNM, although we do not use this nomenclature because in some cases
the IDG is highly pressurized, and thus warm rather than cold. On the
other hand, the HDG can be identified with the ``molecular'' component, 
although we will maintain the notation HDG because we
do not follow the chemistry nor have cooling 
appropriate for the molecular gas. The times shown for figs.\
\ref{fig:P_hists}a and b  are
the same as the eralier times in figures \ref{fig:rho_hists}a and b.
Figures \ref{fig:P_vs_rho}a and
b in turn show the distribution of points in
the simulations in the $(P,\rho)$ plane, overlayed on the $\Peq$ vs.\
$\rho$ curve, at the same times as fig.\ \ref{fig:P_hists}.

Several points are worth noting in these figures. First, again the
pressure of the unperturbed inflow gas is noticeable as the sharp peak
at $\log P = 3.38$ ($P=2400$ K cm$^{-3}$) in the pressure histograms of
the entire simulations. However, the global shape of the two histograms
is very different. The total histogram for run M1.2L32 (fig.\
\ref{fig:P_hists}a) is quite narrow, with a total width of slightly over
one order of magnitude, and moreover has a second, wide maximum
centered at $P \sim 5000$ K cm$^{-3}$. Instead, the
total pressure histogram of run M2.4L16 (fig.\ \ref{fig:P_hists}b)
extends over two orders of magnitude and has a nearly lognormal
shape (except for the sharp peak noted above), rather than the bimodal
shape of the $\Mr =1.2$ run. All of this indicates a more
developed state of the turbulence in run M2.4L16.

Focusing on the pressure distributions for the IDG and HDG, we note that,
in run M1.2L32, the pressure in the IDG is mostly confined to lower
values than 
those of the HDG, in the range 1000--4000 K cm$^{-3}$, with a very low
tail extending to $\sim 7000$ K$\pcc$. The
most probable value of the pressure of this gas practically coincides 
with that of the unperturbed inflowing WNM, with $P \sim 2400$ K
cm$^{-3}$. The HDG, on the other hand, is systematically
overpressured with respect to the IDG and the unperturbed WNM, with
$4000 \lesssim P/({\rm K~cm}^{-3}) \lesssim$ 10,000. It is interesting
that the broad high-pressure maximum in the total histogram overlaps
with the range of pressure values of the ``molecular'' gas. This
maximum corresponds to the shocked, low-density gas that is in
transit from 
the warm to the cold phases, crossing the unstable density range, $0.6
\lesssim n \lesssim 7 \pcc$, as can be seen in fig.\
\ref{fig:P_vs_rho}a. \emph{The pressure coincidence between the
HDG and the shocked unstable gas strongly
suggests that the HDG is in pressure balance with the shocked,
compressed gas, rather than with the ambient WNM, explaining its
higher-than-average pressure.}

In the case of run M2.4L16, some
new features arise. Most notably, the pressure
distribution of the WNM and the IDG now extend beyond that
of the HDG (see also fig.\ \ref{fig:P_vs_rho}b). This 
is somewhat surprising, and probably indicates that a substantial
fraction of the pressure in the shocked gas is converted into kinetic
energy of the HDG by the dynamical instabilities, rather than into
internal energy. This picture is supported by the fact discussed in the
next subsection, that the ratio of turbulent kinetic-to-internal energy
density is highest in the HDG in this high-$\Mr$ run. It is also interesting
that the pressure distribution of the HDG extends over a very similar
range ($4000 \lesssim P/({\rm K~cm}^{-3}) \lesssim 10000$) as that of the
mildly supersonic case M1.2L32, in spite of the much higher pressures
present in the IDG and WNM distributions.

\subsubsubsection{Energy densities and rms speeds in the various regimes}

Table \ref{tab:energies} summarizes the energy densities of the
inflowing WNM, the IDG and the HDG for the two runs under
consideration. For the IDG and the HDG, the table additionally gives the
rms speed, rms Mach number, and mean temperature.
The data for each run are given at two times: the
time at which the shocks first touch the $x$-boundaries, and the (later)
time at which the density and pressure histograms become stationary
(cf.\ \S \ref{sec:global_evol}). We see that the statistics for the IDG
indiceate that indeed it is slightly more turbulent at the later times
for both runs, with larger velocity dispersions, rms Mach numbers and
lower mean temperatures (indicating larger densities due to stronger
compressions). The statistics for the HDG, however, are nearly
indistinguishable at the two times.

Note that the velocity statistics
reported in Table \ref{tab:energies} refer to the total 
velocity dispersion of these components, and thus includes the bulk
motions of the moving dense gas parcels. Although we have not measured
it here, it has consistently 
been reported by various groups \citep{KI02,Heitsch_etal05} that the
\emph{internal} velocity dispersion of the dense gas regions is
subsonic. We do not attempt these measurments here, and defer
the task for future papers using higher resolution simulations.

It can be seen from Table \ref{tab:energies} that in run M1.2L32 both
the IDG and the 
HDG have higher internal than kinetic energy densities. The
opposite is true for run M2.4L16, in which the kinetic 
energy density in these two components is 2.5--3.5 times larger than the
internal energy density. Note in fact that the
kinetic energy density in the turbulent IDG and HDG is 
larger than that in the inflowing streams. This does not
constitute any violation of energy conservation, since the volume
occupied by the IDG and the HDG is small compared with that of the
diffuse WNM. 

Finally, from the rms speed and Mach number data for
the IDG and HDG in the two runs we see that their motions are in general
transonic (in the IDG) or supersonic (in the HDG) 
with respect to their own sound speeds. The rms Mach number in the HDG is
lower than typical values for molecular clouds, but this can be
attributed mainly to the relatively low resolution of the simulations,
causing the velocities and the density fluctuations to be somewhat
damped at the scales of the dense gas by numerical diffusion. This
causes lower velocities and higher temperatures, thus lowering the Mach
number. Moreover, we do not model the transition
from atomic to molecular hydrogen, so that the simulations do not
account for the reduction of the sound speed upon the formation of
molecules. But we see that the velocity dispersion, $\sim 1.7~\kms$
would correspond to Mach numbers $\sim 8.5$ in molecular gas at $T \sim
10$ K.

\section{Discussion and comparison with previous work} \label{sec:discussion}

\subsection{CNM sheet formation at early stages} \label{sec:disc_early}

Our results from both the qualitative analysis and the numerical
simulations show that during the early stages of evolution, the
collision of WNM streams may form thin sheets of CNM, whose properties
are described in \S\S \ref{sec:early} and \ref{sec:mod_sim_comp} and
fig.\ \ref{fig:ncold_Minfl}. Since the sheets last the longest
at low inflow Mach numbers, we consider the results of simulation
M1.03L64-1Dhr, reported in \S \ref{sec:mod_sim_comp}. 

According to this simulation, the outward velocity of the cold layer
boundary is $\vs \sim 0.015$ km s$^{-1}$, implying
that the layer thickness is
\begin{equation}
\ell = 0.0306  \left[\frac{t}{1 {\rm ~Myr}}\right] {\rm ~pc},
\end{equation}
and, with a number density $n_3 \sim 255 \pcc$, its column density is 
\begin{equation}
N_3 = 2.4\times 10^{19} \left[\frac{t}{1 {\rm ~Myr}}\right] \psc.
\end{equation}
The pressure in the cold layer is
$P_3=6650$ K $\pcc$, and therefore its temperature is $T \approx 26$
K. These values are interestingly similar to those derived by \citet[][
hereafter HT03]{HT03} \citep[see also][]{Heiles04} for cloud ``A'' of 
\citet{KV72}, of $N \sim 0.2 \times 10^{20} \psc$, $T \sim 20$ K, $n
\sim 150 \pcc$ and thickness $\sim 0.05$ pc. In general, the column
densities in fig.\ 
\ref{fig:ncold_Minfl} are within the range of
the values reported in Table 5 of HT03. These similarities
suggest that sheets such as those reported by HT03 can be formed
by transonic compressions in the WNM, as modeled by our simulations.

Two important remarks are in order. First, \citet{Heiles04} assigns to
this cloud a 
characteristic time scale of $\sim 5 \times 10^4$ yr, on the basis of
an observed line-of-sight turbulent velocity component $\sim 1$ km
s$^{-1}$ and the thickness of 0.05 pc, while our estimates above are for
a time of 1 Myr after the compression started. This apparent discrepancy
may be resolved as follows. From our fig.\ \ref{fig:Mach0.96_interm} we
see that, although the velocity at the center of the slab is very
close to zero, it rapidly increases in the transition front, since the
velocity of the 
gas right outside the cold layer is close to $3 \kms$. Thus, sampling
the gas out to sufficiently distant positions from the collision center
should pick up higher velocities. We can investigate this effect in the
1D simulations, which by construction are not turbulent, to determine
the linewidths that can be produced by the inflow alone.

\begin{figure}
\epsscale{1.}
\plotone{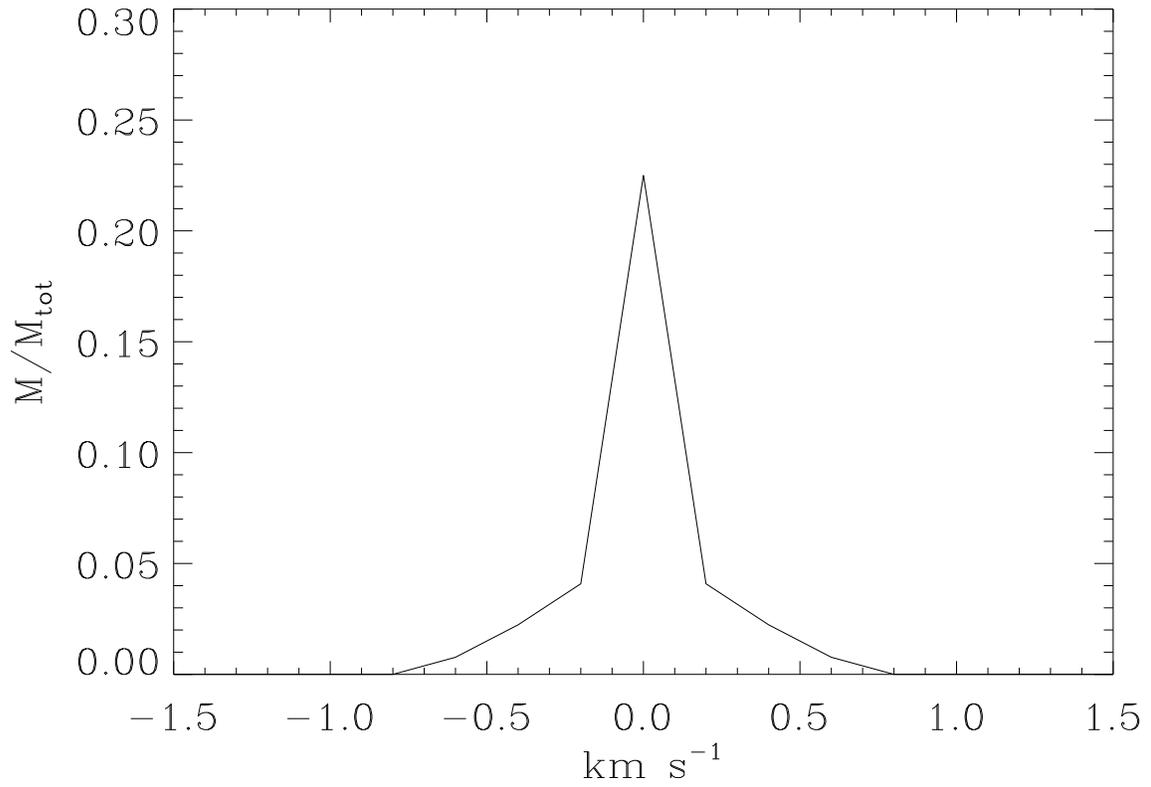}
\caption{Mass-weighted velocity histogram (simulated line-profile) of
the gas in run M1.03L641Dhr at temperatures $T < 500$ K. The histogram
has a velocity resolution of 0.2$\kms$. A central line of width $\sim 0.5
\kms$ and broad wings with FWHM of $\sim 1 \kms$ are observed.}
\label{fig:M1.03_rho_v_hist}
\end{figure}

Our simulations do not yet
resolve well the cold slab at 1 Myr, although run M1.03L64-1Dhr begins to 
resolve it at $t= 5.3$ Myr (cf.\ \S \ref{sec:num_res}). The line profile
of the slab at this time can be approximated by the mass-weighted 
velocity histogram for gas with $T<500$ K, using 
a velocity resolution of $0.2 \kms$ in order to crudely mimic blurring
by thermal broadening. The resulting profile is shown in fig.\
\ref{fig:M1.03_rho_v_hist}. A central line with FWHM $\sim 0.5 \kms$ is
seen, with broad wings of width $\sim 1 \kms$, in reasonable
agreement with the observations. Moreover, we have found that the
linewidth is nearly invariant over time, as a consequence of the fact
that the greatest contribution comes from the material in the interface
between the dense slab and its surrounding medium. Thus, we expect the
line profile at $t=5.3$ Myr to be a good estimate of that at $t=1$ Myr.
These results suggest that the observed linewidths of the thin CNM
sheets can be almost entirely accounted for by the accretion onto the
sheet. Moreover, since none of our 3D simulations develop turbulence by
times as early as $\sim 1$ Myr, these results suggest that the observed
linewidths are not representative of internal motions, but of the
accretion onto the sheets.

The second remark is that HT03 estimated the number density in the sheets 
they observed from the spin temperature, assuming a pressure $P=2250$
K$\pcc$. Our model and simulations both suggest that the sheets are
actually at a significantly higher pressure than the mean interstellar
value, due first to the outer shock and then to the deceleration towards
the center. This would raise their density estimates by factors $\sim
3$, in agreement with the fact that our densities are in general larger,
although making the observed sheets even thinner.  

If our identification of the cold layers in our simulations with the
thin CNM sheets reported by HT03 is correct, then this implies that they
can be the ``little sisters'' of molecular clouds, produced by
compressions that are not strong enough to rapidly develop turbulence
nor produce very dense gas, with the only difference in interpretation
with respect to \citet{Heiles04} being
that the 1-km s$^{-1}$ linewidth does not imply rapid destruction of the
sheet, but instead just represents the
velocity of the gas entering the sheet. The appropriate
destruction time is that required for the \emph{development} of
turbulence in the cold layer which, as we have seen, is a rapidly
varying function 
of the inflow Mach number, but is in general greater than 5 Myr in the
cases we have investigated.

\subsection{Late stages and turbulence} \label{sec:disc_late}

The results of \S \ref{sec:late} are complementary to those presented by
\citet{KI02} and \citet{Heitsch_etal05}. 
In particular, the latter authors presented a physical setup
very similar to ours, at higher resolutions in
two-dimensions, and recognized three instabilities that may be at
play in the problem, namely TI, NTSI and also the Kelvin-Helmholz
instability, with the former one working to create the dense cold layer and
the latter two working to produce disordered motions. In their study,
these authors focused on the competition between these instabilities,
the mass distribution of the cold clumps, and the generation of
vorticity in the cold gas. In our study, we have focused primarily on
the pressure of the cold gas, as a step in understanding the
overpressured nature of molecular clouds, the fractions of thermal and
kinetic energies in the cold gas,  and the rapidity of turbulence
development as a function of the inflow Mach number.

In summary, our work, taken together with those previous studies
strongly suggests that various physical properties of molecular clouds,
such as rms velocities of a few $\kms$, densities of several hundred
$\pcc$, and thermal pressures several times larger than the mean
interstellar values, can be produced during the formation stages of the
clouds, without the need for external energy sources, other than the
ones that produced the large-scale compression.

\section{Summary and conclusions} \label{sec:concl}

In this paper we have studied the process of cloud formation by
large-scale stream collisions in the WNM, presenting a simple
analytical study of the initial stages, and numerical
simulations of the whole process. The
analytical model and high-resolution 1D simulations show that thin
sheets of cold neutral medium can be formed within the
shock-bounded layer by transonic compressions ($\Mr \sim 1$) on
timescales $\sim 1$ Myr. These sheets are reminiscent 
of those reported by \citet{HT03}, with column densities $\sim 2.5
\times 10^{19} \psc$, thicknesses $\sim 0.03$ pc, temperatures $\sim 25$
K and pressures $\sim 6500$ K$\pcc$. In our simulations the sheets have
linewidths $\sim 1 \kms$, 
again comparable to the value reported by \citet{Heiles04}, although
these linewidths do not correspond to turbulent motions in the layer,
but rather to the inflowing speed of the gas. Also, our sheets are at
higher pressures than those assumed by \citet{HT03}, implying that their
number densities are higher than those authors estimated.

At later times, the simulations show that the boundary of the cold layer
becomes dynamically unstable, through an NTS-like instability that
occurs even though the flow is always subsonic inside the shocked
layer. Eventually, fully developed turbulence arises, on times that can
be as short as $\sim 5$ Myr for inflow Mach numbers $\Mr=2.4$, and as
long as over 80 Myr for $\Mr =1.03$. In this turbulent regime, the
highest-density gas (HDG, with $n > 100 \pcc$) is always overpressured
with respect to the mean WNM 
pressure by factors 1.5--5. Since our simulations do not include
self-gravity, this result shows that dense, overpressured gas can be
readily formed by dynamical compressions in the WNM, possibly explaining
at least part of the excess pressure in molecular clouds. The
intermediate-density gas (IDG, with $10 < n [{\rm cm}^{-3}] < 100$) has
a significant pressure scatter at a given value of the density, which
increases with inflow Mach number, so that at $\Mr =2.4$ a significant
fraction of the IDG has pressures larger than those of the HDG. In general, the
ratio of internal to kinetic energy density of the inflowing gas changes
as the gas is incorporated  into
the IDG and the HDG, with a tendency to increase as
one considers higher-density gas in the fully turbulent regime. Finally,
the density probability distribution tends to lose the 
bimodal signature of thermal bistability as the inflow Mach number is
increased.

Our calculations are not free of caveats, with the most notable ones being 
our neglect of molecular cooling, thermal conduction, magnetic fields,
and self-gravity. The relatively low resolutions we have used imply that
the structure within the dense gas is not resolved. Finally, due to
somewhat small box sizes used as a compromise between acceptable
resolution at the early times and sufficient spatial coverage at the
late, turbulent times, the simulations cease to be valid in a strict
sense (because the bounding shocks leave the box) before the turbulence
becomes stationary, although their statistical properties at later times
do not seem to be affected by this fact. We plan to address these
shortcomings in future papers. 

Our results, together with those of previous groups
\citep{KI02,Heitsch_etal05} suggest that the turbulence and at least
part of the excess pressure in  molecular clouds are generated during
the compression that forms the clouds themselves, and that the CNM
sheets reported by \citet{HT03} may be formed by the same mechanism,
in cases where the compressions are only mildly supersonic.

\acknowledgments
We are glad to acknowledge useful discussions with Carl Heiles and
Ethan Vishniac, and a useful and thorough report by an anonymous referee. 
This work has received partial support from CONACYT grant 36571-E to E.\
V.-S., Korea Research Foundation grant KRF-2004-015-C00213 to D.\ R., and
from the French national program PCMI to T.\ P. The numerical 
simulations have been performed on the linux cluster at Centro de
Radioastronom\'ia y Astrof\'isica of UNAM, funded by the above CONACYT
grant. We have made 
extensive use of NASA's ADS and LANL's astro-ph abstract services.


\begin{deluxetable}{cccccccccccc}
\rotate
\tablecaption{Numerical simulation parameters.
\label{tab:runs}}
\tablewidth{0pt}
\tablehead{
\colhead{Run name} & 
\colhead{Dimensionality} & 
\colhead{$n_{x,y}$\tablenotemark{a}}  &
\colhead{$n_{z}$\tablenotemark{b}}  &
\colhead{$\Mr$\tablenotemark{c}}   &
\colhead{$v_1$\tablenotemark{d}}   &
\colhead{$L_0 [{\rm pc}]$\tablenotemark{e}}  & 
\colhead{$t_0 [{\rm Myr}]$\tablenotemark{f}}  &
\colhead{$\Delta x$ [pc]\tablenotemark{g}} & 
$\Delta t$ [Myr]\tablenotemark{h} 
}
\startdata
%
%
M2.4L16		& 3D & 200 & 200 & 2.4 & 23.7 & 16 & $1.33 $ & 0.08 & 0.67\\
%
%
%
%
M1.2L32 	& 3D & 200 & 200& 1.2  & 11.9 & 32 & $2.66 $ & 0.16 & 1.33\\
M1.2L32-1D 	& 1D & 1000 & --- & 1.2 & 11.9 & 32 & $2.66 $ & 0.032 & ---\\
M1.03L64	& 3D & 200 & 200& 1.03 & 10.2 & 64 & $5.33 $ & 0.32 & 2.67\\
M1.03L64-LZ	& 3D & 200 & 50	& 1.03 & 10.2 & 64 & $5.33 $ & 0.32 & ---\\
M1.03L64-hr-LZ	& 3D & 400 & 50	& 1.03 & 10.2 & 64 & $5.33 $ & 0.16 & ---\\
M1.03L64-1D	& 1D & 1000 & --- & 1.03 & 10.2 & 64 & $5.33 $ & 0.064 & ---\\
M1.03L64-1D-hr	& 1D & 4000 & --- & 1.03 & 10.2 & 64 & $5.33 $ & 0.016 & ---\\

\enddata




\tablenotetext{a} {Resolution in the $x$ (and $y$) direction(s) (in 3D).}
\tablenotetext{b} {Resolution in the $z$ direction.}
\tablenotetext{c} {Inflow Mach number.}
\tablenotetext{d} {Inflow speed in simulation frame.}
\tablenotetext{e} {Physical box size in parsecs.}
\tablenotetext{f} {Physical time unit.}
\tablenotetext{g} {Minimum resolved scale.}
\tablenotetext{h} {Time interval between frames in animation.}


\end{deluxetable}

\begin{deluxetable}{cccccccccccc}
\rotate
\tablecaption{Physical conditions in the turbulent gas.
\label{tab:energies}}
\tablewidth{0pt}
\tablehead{
\colhead{Run name(@ time)} & Component &
\colhead{$e_{\rm th}$\tablenotemark{a} [erg $\pcc$]}   & \colhead{$e_{\rm
k}$\tablenotemark{b} [erg $\pcc$]}  & \colhead{$u_{\rm
rms}$\tablenotemark{c} [km s$^{-1}$]} 
&  \colhead{$M_{\rm rms}$\tablenotemark{d}} & $T_{\rm mean}$ [K]
\tablenotemark{e} 
}
\startdata
M1.2L32 & Inflow & $5.0 \times 10^{-13}$ & $3.8 \times 10^{-13}$\\
(@ 42.6 Myr) & IDG & $5.6 \times 10^{-13}$ & $4.7 \times 10^{-13}$ & 1.3 & 1.25 & 45. \\
	& HDG	 &$ 1.0 \times 10^{-12}$ & $ 6.0 \times 10^{-13}$ & 0.72 & 1.0 & 21. \\
\\
(@ 47.9 Myr) & IDG & $5.4 \times 10^{-13}$ & $ 5.2 \times 10^{-13}$ & 1.4 & 1.3 & 45. \\
	& HDG	 &$ 1.0 \times 10^{-12}$ & $ 6.4 \times 10^{-13}$ & 0.73 & 1.1 & 21. \\
\\
\\
M2.4L16 & Inflow & $5.0 \times 10^{-13}$ & $1.5 \times 10^{-12}$\\
(@ 7.37 Myr) & IDG & $7.8 \times 10^{-13}$ & $2.0 \times 10^{-12}$ & 2.8 & 2.3 & 92. \\
	& HDG	 &$ 1.1 \times 10^{-12}$ & $3.6 \times 10^{-12}$ & 1.7 & 2.4 & 21. \\
\\
(@ 10.7 Myr) & IDG & $6.8 \times 10^{-13}$ & $2.4 \times 10^{-12}$ & 3.1 & 2.6 & 78. \\
	& HDG	 &$ 1.1 \times 10^{-12}$ & $3.6 \times 10^{-12}$ & 1.7 & 2.4 & 21. \\
\enddata

\tablenotetext{a} {Mean internal energy density in component.}
\tablenotetext{b} {Mean turbulent kinetic energy density in component.}
\tablenotetext{c} {\emph{rms} speed in component.}
\tablenotetext{d} {\emph{rms} Mach number in component.}
\tablenotetext{e} {Mean temperature in component}
\tablenotetext{f} {Inflowing WNM, at $n=0.338 \pcc$,  $T=7100$ K and
Mach number indicated by the run name.}
\tablenotetext{g} {Intermediate-density gas ($10 < n~[{\rm cm}^{-3}] <
100$).}
\tablenotetext{h} {High-density gas ($ n > 100 \pcc$).}


\end{deluxetable}

\clearpage

\clearpage

\end{document}